\documentclass[12pt,preprint]{aastex}

\usepackage{graphicx,amsmath,color,natbib,soul,morefloats,longtable}
\bibpunct{(}{)}{;}{a}{}{,}

\shorttitle{Substructure in the MW Stellar Halo}

\shortauthors{Xue et al.}

\begin{document}

\title{Quantifying Kinematic Substructure in the Milky Way's Stellar Halo}

\author{Xiang-Xiang Xue\altaffilmark{1,2}, Hans-Walter Rix\altaffilmark{2},
  Brian Yanny\altaffilmark{3}, Timothy C. Beers\altaffilmark{4}, Eric F.
  Bell\altaffilmark{5,2}, Gang Zhao\altaffilmark{1}, James S.
  Bullock\altaffilmark{6}, Kathryn V. Johnston\altaffilmark{7}, Heather
  Morrison\altaffilmark{8}, Constance Rockosi\altaffilmark{9}, Sergey
  E. Koposov\altaffilmark{10,2,11}, Xi Kang\altaffilmark{12,2}, Chao
  Liu\altaffilmark{2}, Ali Luo\altaffilmark{1}, Young Sun Lee\altaffilmark{4},
  Benjamin. A. Weaver\altaffilmark{13}}

\altaffiltext{1}{Key Lab of Optical Astronomy, National Astronomical
  Observatories, CAS, 20A Datun Road, Chaoyang District, 100012, Beijing,
  China}

\altaffiltext{2}{Max-Planck-Institute for Astronomy K\"{o}nigstuhl 17,
  D-69117, Heidelberg, Germany}

\altaffiltext{3}{Fermi National Accelerator Laboratory, P.O.  Box 500 Batavia,
  IL 60510-5011, USA}

\altaffiltext{4}{Department of Physics and Astronomy and JINA: Joint Institute
  for Nuclear Astrophysics, Michigan State University, E. Lansing, MI 48824,
  USA}

\altaffiltext{5}{Department of Astronomy, University of Michigan, 500 Church
  Street, Ann Arbor, Michigan, 48109, USA}

\altaffiltext{6}{Center for Cosmology, Department of Physics and Astronomy,
  University of California, Irvine, CA 92697, USA}

\altaffiltext{7}{Astronomy Department, Columbia University, New York, NY
  10027, USA}

\altaffiltext{8}{Department of Astronomy, Case Western Reserve University,
  Cleveland, OH 44106, USA}

\altaffiltext{9}{Lick Observatory/University of California, Santa Cruz, CA
  95060, USA} 

\altaffiltext{10}{Institute of Astronomy, Madingley Road,
  Cambridge CB3 0HA, UK}

\altaffiltext{11}{Sternberg Astronomical Institute, Universitetskiy pr. 13,
  119992 Moscow, Russia}

\altaffiltext{12}{Purple Mountain Observatory, CAS, 2 West Beijing Road,
  Nanjing, 210008, China}

\altaffiltext{13}{Center for Cosmology and Particle Physics, New
York University, New York, NY 10003, USA}

\begin{abstract}

We present and analyze the positions, distances, and radial velocities for
over 4000 blue horizontal-branch (BHB) stars in the Milky Way's halo, drawn
from SDSS DR8. We search for position-velocity substructure in these data, a
signature of the hierarchical assembly of the stellar halo. Using a cumulative
``close pair distribution'' (CPD) as a statistic in the 4-dimensional space of
sky position, distance, and velocity, we quantify the presence of
position-velocity substructure at high statistical significance among the BHB
stars: pairs of BHB stars that are close in position on the sky tend to have
more similar distances and radial velocities compared to a random sampling of
these overall distributions. We make analogous mock-observations of 11
numerical halo formation simulations, in which the stellar halo is entirely
composed of disrupted satellite debris, and find a level of substructure
comparable to that seen in the actually observed BHB star sample. This result
quantitatively confirms the hierarchical build-up of the stellar halo through
a signature in phase (position-velocity) space. In detail, the structure
present in the BHB stars is somewhat less prominent than that seen in most
simulated halos, quite possibly because BHB stars represent an older
sub-population. BHB stars located beyond 20 kpc from the Galactic center
exhibit stronger substructure than at $\rm r_{gc} < 20$ kpc.

\end{abstract}

\keywords{galaxies: individual(Milky Way) ---  Galaxy: halo --- Galaxy: structure --- stars: horizontal-branch ---
  stars: kinematics and dynamics}


\section{Introduction}

The current hierarchical structure formation paradigm implies that the
formation of our Milky Way entailed a sequence of dark matter driven accretion
and merger events \citep{Searle1978,White1978,Blumenthal1984}. This naturally
results in the expectation that the stellar halo should be largely built up
from stars of tidally disrupted satellite galaxies, resulting in substructure
that may appear as stellar streams with different degrees of phase-mixing
\citep[e.g.][hereafter BJ05]{Bullock2001,Cooper2010,Bullock2005}. Because stars are gravitationally collisionless systems, their
phase-space (spatial and velocity) distributions encode and retain aspects of
their origin. This implies that an analysis of substructure in the
position-velocity distribution of stars in the halo is a direct test for
hierarchical models of galaxy formation.

In the past decades, observational evidence of spatial substructure has indeed
been found in the Milky Way, both near the Sun \citep{Majewski1996, Helmi1999}
and at larger distances \citep{Ibata1994, Ibata1995}. The most prominent
example is the discovery of the Sagittarius dwarf galaxy \citep{Ibata1994,
  Ibata1995, Yanny2000} and its trails of debris \citep{Ibata2001b,
  Majewski2003}.

In nearby samples of stars, where the full 6D phase-space coordinates can be
measured, substructure in the stellar distribution is seen in velocity space,
or even in the integrals of motion \citep{Dehnen1998, Helmi1999, Klement2008,
  Klement2009, Morrison2009, Smith2009}. At distances from the Sun
characteristic of the stellar halo, $\sim$ 20 kpc, individual transverse
velocities are all but impossible to measure from proper motions with current
technology. The available observables are therefore the position in the sky, a
distance estimate from photometric or spectroscopic luminosity determinations,
and line-of-sight velocity: $\alpha, \delta,d,$ and $V_{los}$. When averaged
over large angular areas and broad distance ranges, the line-of-sight
kinematics of the Milky Way halo stars at 10-50 kpc are well-described by a
simple Gaussian with $\rm \sigma_{los}~\approx~111~km~s^{-1}$
\citep[][hereafter X08]{Xue2008}. However, because the stellar halo is
collisionless, preserving phase-space density, substructure in position space
necessarily implies substructure in velocity space. Recent work by
\citet{Starkenburg2009} and \citet{dePropris2010} indicate that the Milky
Way's stellar halo indeed possesses detectable position-velocity substructure.
\citet{Schlaufman2009} have shown that metal-poor halo stars within $\rm
\sim~17.5$ kpc from the Sun exhibit clear evidence for velocity clustering on
very small spatial scales (which the authors refer to as Elements of Cold HalO
Substructure, or ECHOS).

With the development of large-scale sky surveys, such as the Two Micron All
Sky Survey \citep[2MASS;][]{Skrutskie2006}, the Sloan Digital Sky Survey
\citep[SDSS;][]{York2000, Stoughton2002, Abazajian2003, Abazajian2004,
  Abazajian2005, Abazajian2009, Adelman-McCarthy2006, Adelman-McCarthy2007,
  Adelman-McCarthy2008}, and the follow-up SEGUE survey \citep{Yanny2009b}, we
have an unprecedented opportunity to examine Milky Way halo streams in detail
\citep{Ivezic2000, Yanny2000, Newberg2002, Majewski2003, Yanny2003,
  Newberg2007, Yanny2009a}. Halo star samples are now of sufficient size and
quality that a direct statistical comparison with models, such as BJ05, has
become possible.

A first quantitative comparison indicated that the observed level of {\it
  spatial} substructure (on all scales) is similar to that expected from those
simulations, where the halo is composed entirely of disrupted satellites
\citep[][hereafter B08]{Bell2008}. Imaging surveys of M31 \citep{Ibata2007}
have revealed a similarly rich set of substructure in the stellar halo of that
galaxy. Based on photometry of main sequence turn-off (MSTO) stars, B08
constructed a coarse 3D map of the stellar halo density, with almost a factor
of two uncertainty in distances. BHB stars are a much rarer tracer of the old
metal-poor population, but have the great advantage of being luminous, with
M$_g ~\sim~ +0.7$ (vs. M$_g~\sim~3.5$ for MSTO stars) and of having precise
distance estimates ($\sim~5\%$; X08). BHB stars have also been a special
spectroscopic target class in SDSS and SEGUE \citep[e.g.,
][]{Yanny2009b}. Hence, the sample of possible BHB stars with spectra from
SDSS constitutes by far the largest set of luminous tracers (extending to
distances of $\sim~80$ kpc) of the Milky Way's stellar halo with available
four dimensional ($\alpha,\delta,d,V_{los}$) information, where the distances
are accurate to 5\% and the radial velocities accurate to $\rm5\sim20$
km~s$^{-1}$. This sample enables the first attempt at checking that the
statistical properties of kinematics matches (or not) model expectations.

This paper describes a large sample of probable BHB stars with measured
kinematics, and presents an exploration of how to quantify
position-velocity substructure in the Milky Way's stellar halo in order to
compare the observation to simulations such as from BJ05.  It is certainly
possible to pick out the kinematic signature of the Sagittarius stream
\citep[e.g., ][]{Ibata2001b, Starkenburg2009} in these data. However, what we
aim for here is to devise a simple objective measure for quantifying such
substructure. Specifically, we employ the close pair distribution (CPD)
statistic, $F=w_{\theta}\theta^2+w_{\Delta d}(\Delta d)^2+w_{\Delta
  V_{los}}(\Delta V_{los})^2$, to detect substructure, following
\citet{Starkenburg2009}.  Here, $\theta,~\Delta d,~\Delta V_{los}$ are the
angular, distance, and velocity separation of pairs of stars, and $w_{\theta},
~w_{\Delta d}$, and $w_{\Delta V_{los}}$ are suitable weights. The idea is
that a structured or ``clumpy'' position-velocity distribution will have more
pairs with small F than a suitably chosen random distribution. As also argued
by B08, it is important for quantitative data-model comparisons to have a
general statistical measure of substructure, rather than specifically
searching for (here, kinematical) substructure associated with a particular
feature, such as the Sagittarius stream, so we also explore what we should
expect from the BJ05 models, and compare with the observations.

This paper is organized as follows. In Section 2 we present the sample of BHB
stars.  Section 3 provides the definition of the close pair distribution (CPD)
as a statistic, and describes its application to the sample of BHB stars.  The
analogous CPD for the BJ05 simulations and their statistical analysis is
presented in Section 4. Conclusions from the comparisons between observations
and simulations are presented in Section 5.

\section{The Spectroscopic Sample of BHB stars from SDSS DR8}

SDSS-I was an imaging and spectroscopic survey that began routine operations
in April 2000, and continued through June 2005. The SDSS and its extensions
are using a dedicated 2.5m telescope \citep{Gunn2006} located at the Apache
Point Observatory in New Mexico. The Sloan Extension for Galactic
Understanding and Exploration (SEGUE) is one of the three key projects (the
legacy survey, the supernova survey, and SEGUE) in the recently completed
first extension of the Sloan Digital Sky Survey, known collectively as
SDSS-II. The SEGUE program, which ran from July 2005 to July 2008, obtained
$ugriz$ imaging of some 3500 deg$^2$ of sky outside of the SDSS-I footprint
\citep{Fukugita1996, Gunn1998, Gunn2006, York2000, Hogg2001, Smith2002,
  Stoughton2002, Abazajian2003, Abazajian2004, Abazajian2005, Abazajian2009,
  Pier2003, Ivezic2004, Adelman-McCarthy2006, Adelman-McCarthy2007,
  Adelman-McCarthy2008, Tucker2006}, with special attention being given to
scans of lower Galactic latitudes ($|b| < 35\degr$) in order to better probe
the disk/halo interface of the Milky Way. SEGUE obtained some 240,000
medium-resolution spectra of stars in the Galaxy, selected to explore the
nature of stellar populations from 0.5 kpc to 100 kpc
\citep{Yanny2009b}. SDSS-III, which is presently underway, has already
completed the sub-survey SEGUE-2, an extension intended to obtain an
additional sample of over 120,000 spectra for distant stars that are likely to
be members of the outer-halo population of the Galaxy. Data from SEGUE-2 has
been distribution as part of the eighth public data release, DR8
\citep{Aihara2011}.

The SEGUE Stellar Parameter Pipeline processes the wavelength- and
flux-calibrated spectra generated by the standard SDSS spectroscopic reduction
pipeline \citep{Stoughton2002}, obtains equivalent widths and/or line indices
for more than 80 atomic or molecular absorption lines, and estimates T$_{\rm
  eff}$, log g, and [Fe/H] through the application of a number of approaches
\citep[see][]{Lee2008a,Lee2008b, AllendePrieto2008, Smolinski2011}.

We construct a sample of BHB stars from SDSS DR8 with spectra in a fashion
very similar to X08. The spectra are used both to classify stars as BHB and to
obtain measured radial velocities.  In essence, we combine an initial color
cut for BHB candidates with Balmer-line profile shape measurements. The S/N of
the spectra affects the precision of Balmer-line profile shape measurements.
Therefore, spectra are only accepted when the fractional variance between the
best-fitting profile and observed Balmer-line profile is $\leqslant 0.1$. The
color cuts that we used in this paper are:

\medskip

\begin{displaymath}
\rm 0.8<u-g<1.5
\end{displaymath}
\begin{displaymath}
\rm -0.5<g-r<0.0
\end{displaymath}
\noindent The Balmer-line profile cuts used are:
\begin{displaymath}
\rm for~the~H\delta~line:~~~~~~~~~~~ D_{0.2} \leqslant 29 ~\rm
\AA,~~~~~~~~~~~~~~ \rm f_m \leqslant 0.35 ~~~~~~~~~~~~~~~~~~~~~~~~~
\end{displaymath}
\begin{displaymath}
\rm for~the~ H\gamma ~line: ~~~0.75 \leqslant c_\gamma \leqslant 1.25,~~~~~\rm
7~\rm \AA \leqslant b_\gamma \leqslant 10.8-26.5\left(c_\gamma-1.08\right)^2 ~
\end{displaymath}
\noindent where $\rm D_{0.2},~f_m, ~c, and ~b$ are the width of the Balmer
line at 20\% below the local continuum, the flux relative to the continuum at
the line core, and the parameters of the S\'{e}rsic profile, $\rm y = 1.0 -
a\rm \exp{\left[-\left(\frac{|\lambda-\lambda_0|}{b}\right)^c\right]}$,
respectively \citep[see][X08]{Sirko2004}.

The following are the Balmer-line profile cuts used in X08:
\begin{displaymath}
\rm for~the~H\delta~line:~~~~~~~~~~~~~~~~\ 17~\rm \AA \leqslant
D_{0.2}\leqslant 28.5~\rm \AA, ~~~~~~~~~~~\ 0.1 \leqslant f_m \leqslant 0.3\
\end{displaymath}
\begin{displaymath}
\rm for~the~H\gamma ~line: ~~~0.75 \leqslant c_\gamma \leqslant 1.25,~~~~~\rm
7~\rm \AA \leqslant b_\gamma \leqslant 10.8-26.5\left(c_\gamma-1.08\right)^2 ~
\end{displaymath}

\begin{figure}
\includegraphics[width=0.5\textwidth]{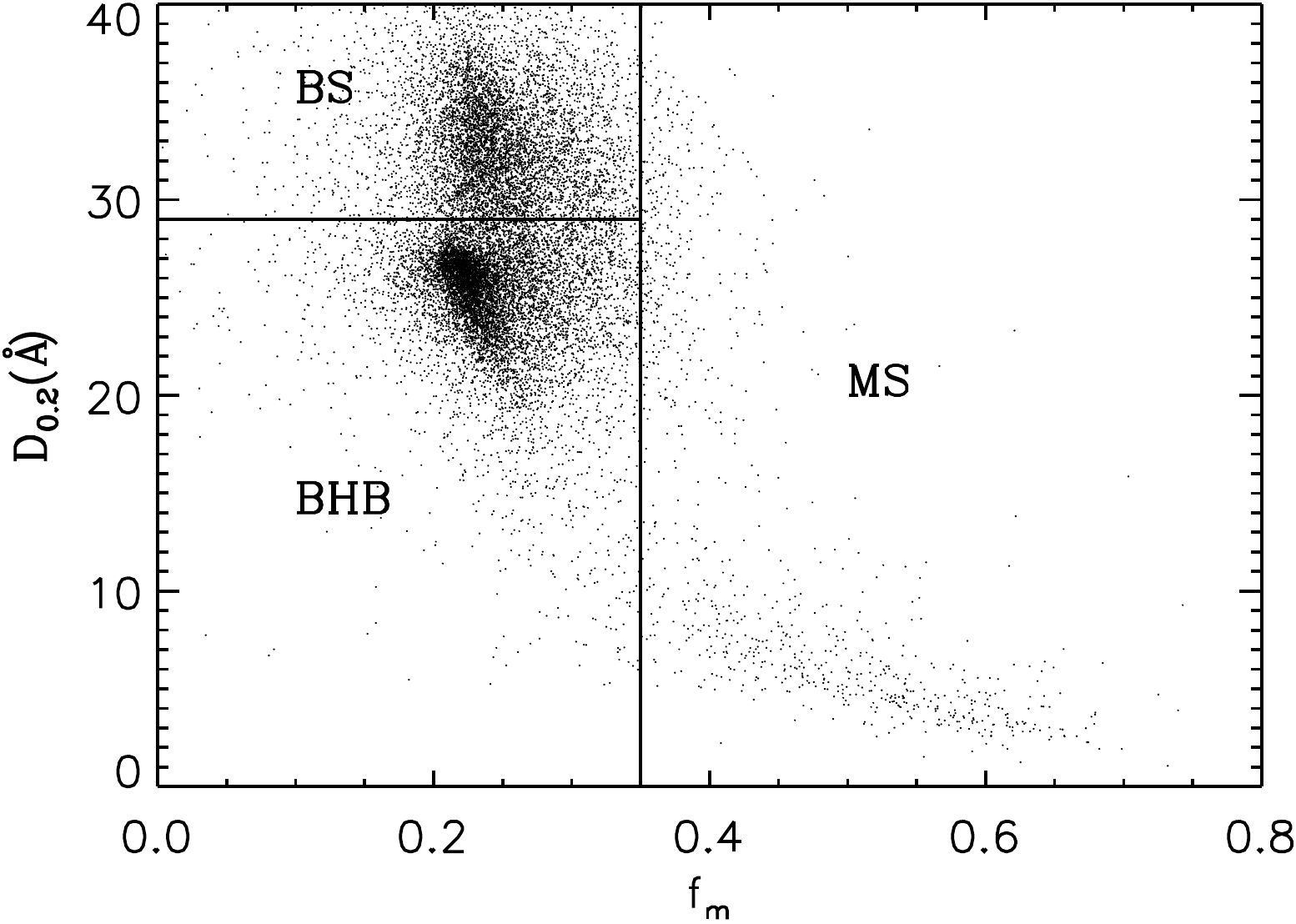}
\includegraphics[width=0.5\textwidth]{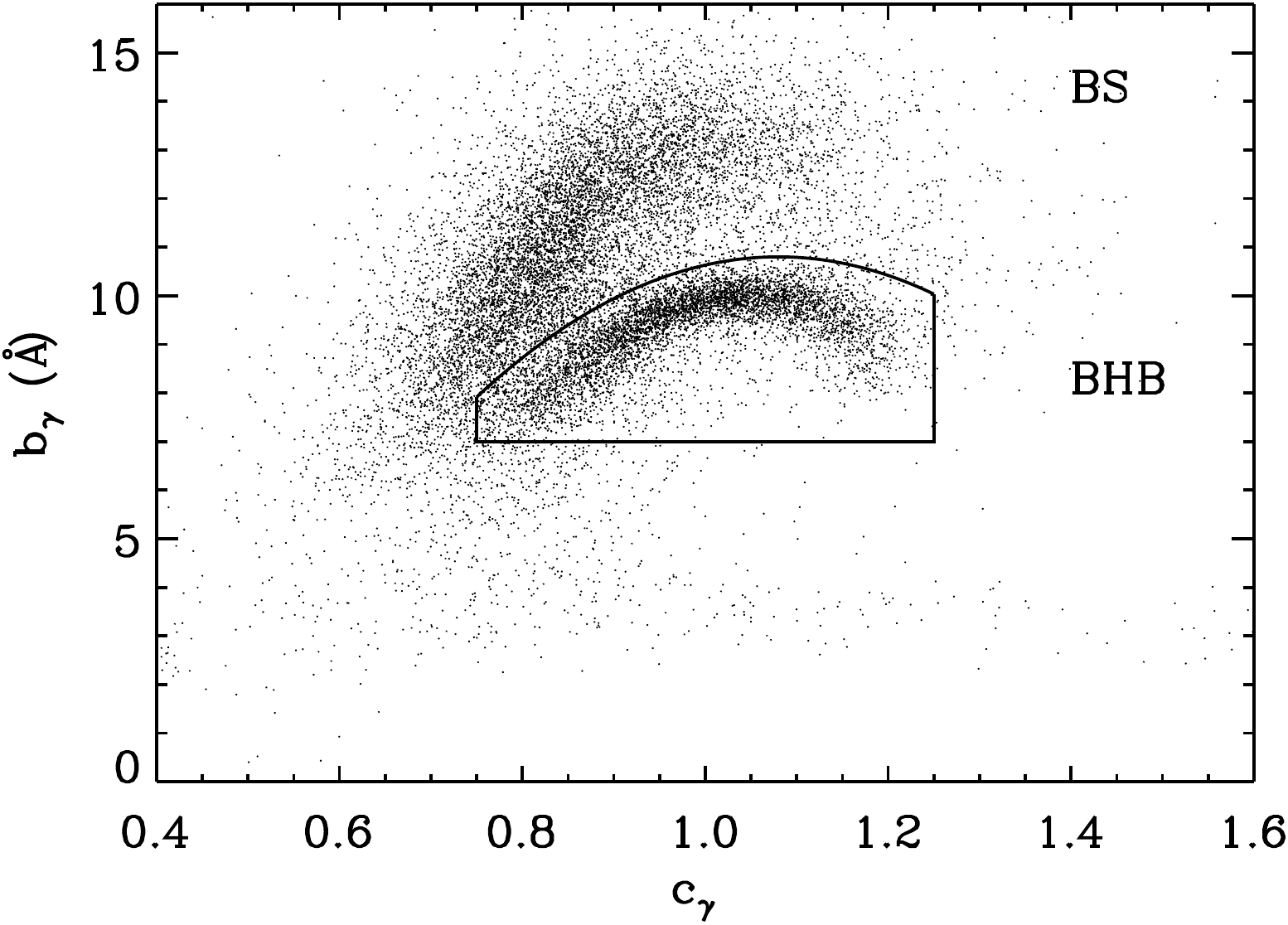}

\caption{BHB star sample selection, based on the Balmer-line shape parameters
  (see Section 2). The left-hand panel shows the $\rm H\delta$ line parameters
  $\rm f_m$ and $\rm D_{0.2}$, divided into three regions (following X08 and
  Sirko et al. 2004): stars with $\rm f_m > 0.35$ are too cool to be BHB stars
  -- they are likely main-sequence stars; the concentration of stars with $\rm
  D_{0.2}>29\rm \AA$ is likely due to blue stragglers (BS) with higher surface
  gravity; the region with $\rm f_m \leqslant 0.35$ and $\rm D_{0.2} \leqslant
  29\rm \AA$ is used as the BHB selection criterion for the $\rm H\delta$,
  $D_{0.2}$, and $f_m$ method. The right-hand panel shows the $\rm
  H\gamma$-line profile parameters $\rm c_\gamma$ and $\rm b_\gamma$ for the
  same stars as in the left panel. Here, BS and BHB stars can be distinguished
  clearly through their bimodal distribution in this plane. The enclosed
  region indicates the $\rm H_\gamma$ scale width-shape criteria that selects
  BHB stars. Our BHB sample is composed of all stars satisfying both criteria
  (left-hand and right-hand panels) simultaneously. This leaves a sample of
  4985 objects with a high probability of proper classification as BHB stars
  (see X08).}
\label{f:f1}
\end{figure}

We retain the color cut \citep{Yanny2000}, but slightly relax the Balmer-line
profile cuts compared to X08, as illustrated in Figure~\ref{f:f1}. Since the
two Balmer-line profile cuts are independent, the relaxed criteria on the
H$\delta$ line should introduce little additional contamination, but
overall it makes the criteria less stringent. As compared with our previous
criteria (X08), the relaxed criteria will have minimal impact on the following
substructure analysis. For further details on the sample selection, we refer
the interested reader to X08 and references therein.

\begin{figure}
\centering
\includegraphics[scale=0.52]{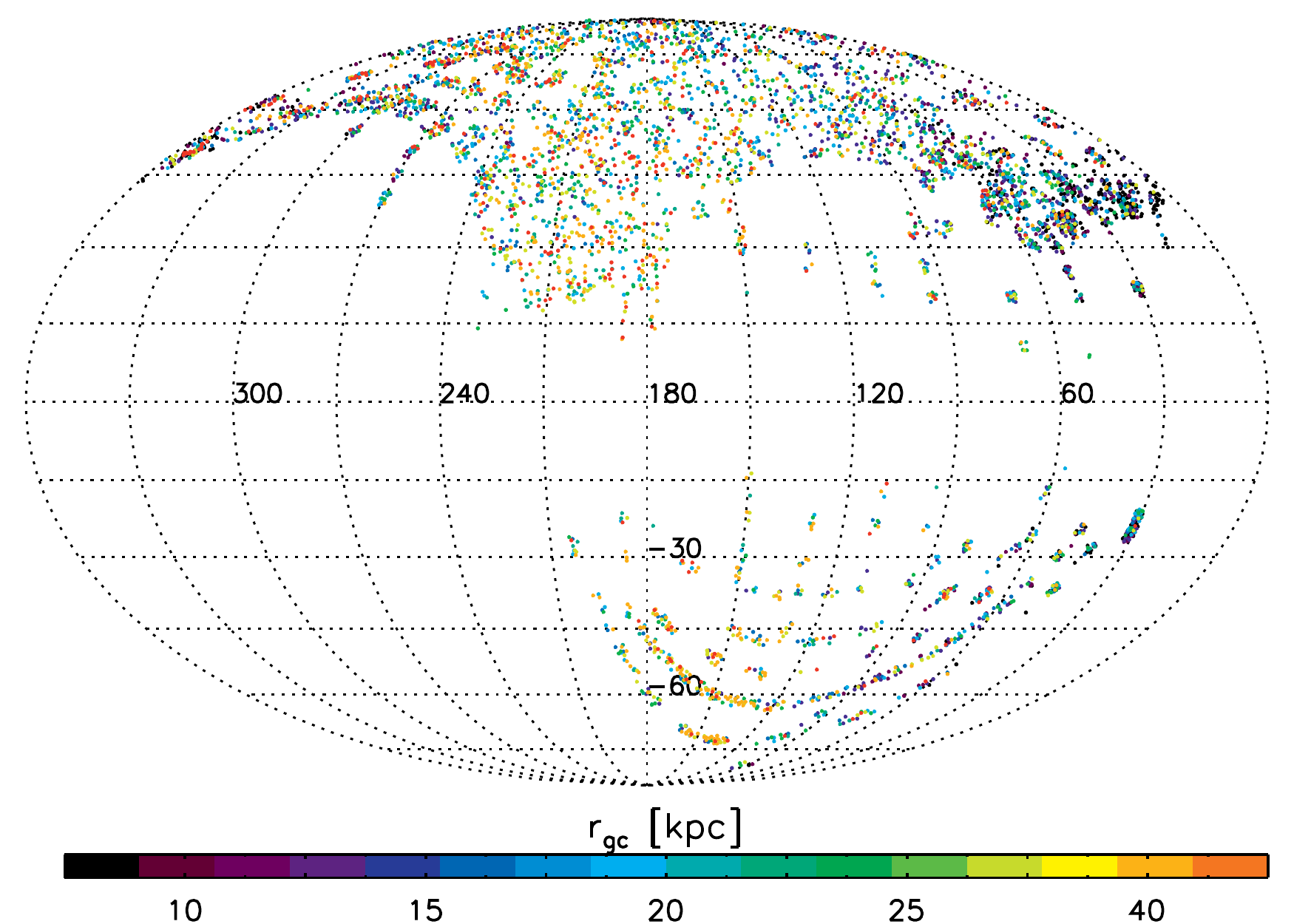}
\includegraphics[scale=0.52]{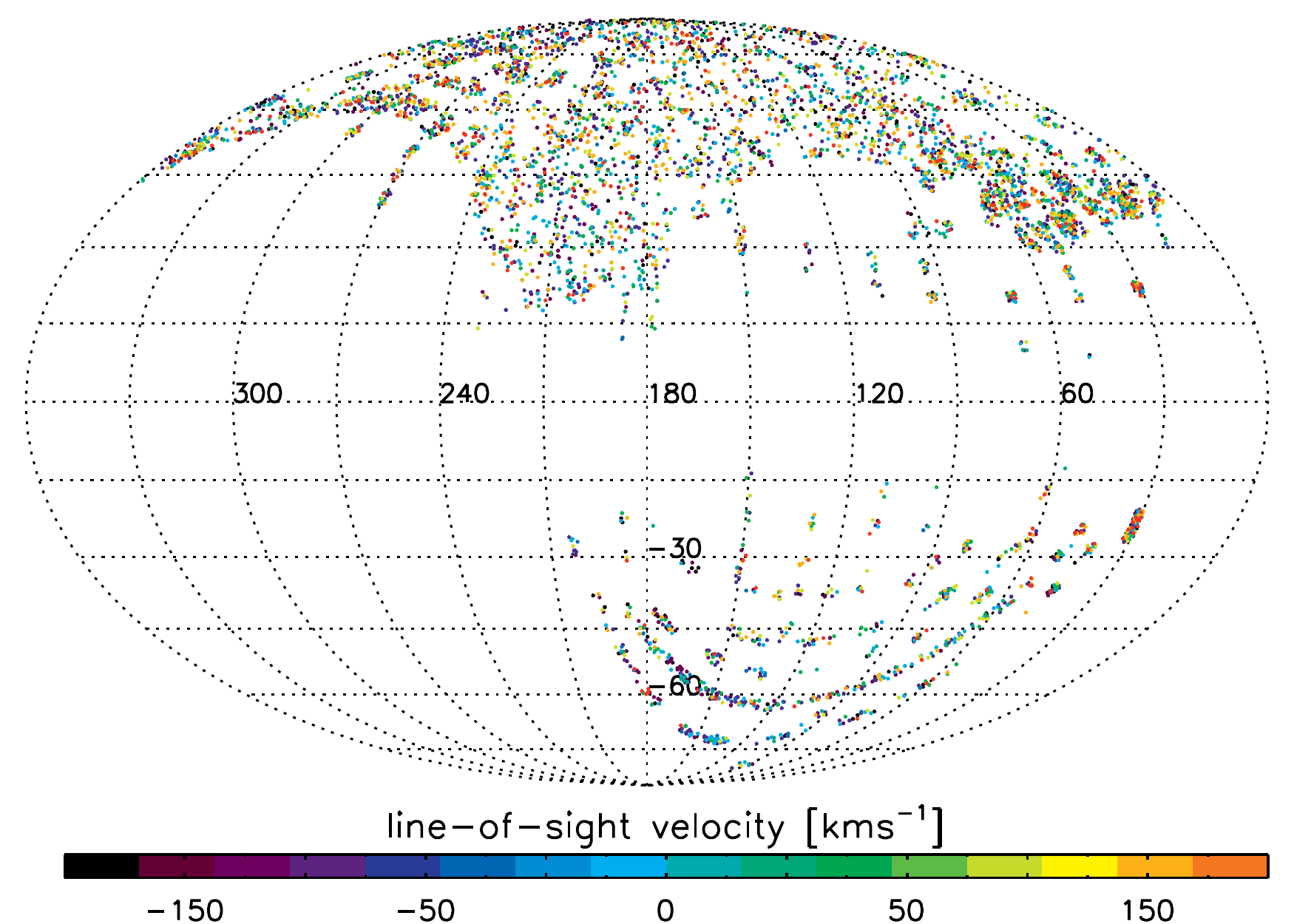}
\includegraphics[scale=0.52]{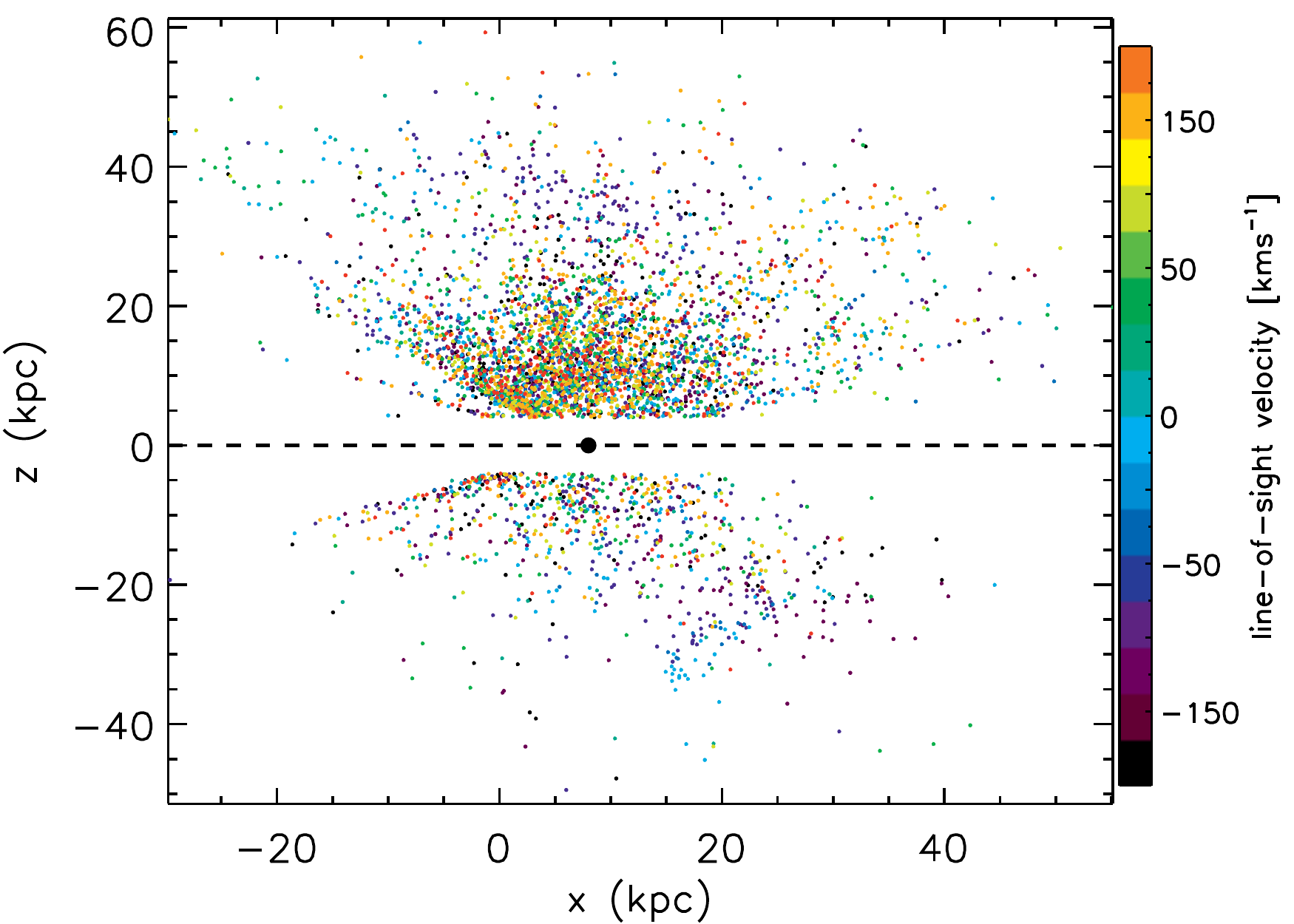}

\caption{Sample properties for the 4625 stars with high probability of being
  BHB (Figure \ref{f:f1}). The first two plots show the Galactic sky coverage
  of the BHB sample, where stars are colored according to radius and
  line-of-sight velocity. The spatial distribution (x-z plane) is shown as the
  third panel. The coordinate system has its origin at the Galactic center;
  the large filled circle on the x-z plot indicates the location of the Sun
  (8.0 kpc, 0 kpc); and the stars are coded according to line-of-sight
  velocity. } \label{f:f2}
\end{figure}

By selecting stars that satisfy the color cuts and both Balmer-line profile
cuts, we obtain a sample of 4985 stars from SDSS DR8 with high BHB
probability (see Table~1 for example), of which there are 4625 halo BHB stars with $\rm |Z|>4$~kpc and
$\rm r_{gc}<60$~kpc and 26 halo BHB stars between $60$ and
$80$~kpc. Figure~\ref{f:f2} shows the sky coverage and spatial distribution of
these 4625 halo BHB stars. Distances were derived from the magnitudes and
colors as in X08. The line-of-sight velocities, $V_{los}$, are converted from
Local Standard of Rest frame to the Galactic Standard of Rest frame by
adopting a value of $220$ km s$^{-1}$ for the Local Standard of Rest (V$_{\rm
  lsr}$) and a solar motion of $(+10.0, ~+5.2, ~+7.2)$ km s$^{-1}$, as in
X08. Small changes that may arise from adopting a different $\rm V_{cir}(R_0)$
pair, e.g., \citet{Bovy2009} or \citet{Koposov2010}, do not matter for the
subsequent analysis.

This sample of halo BHB stars has radial velocity errors of 5-20 km s$^{-1}$
and much more accurate distances than other distant halo stars with available
kinematic information. For instance, distances are $\sim 4\times$ more
accurate than in the sample of halo giants recently used by
\citet{Starkenburg2009} in a search for distant halo substructure, and our
sample is 50-fold larger. \citet{Schlaufman2009} discussed a sample of $\sim
10,000$ metal-poor main sequence turnoff (MPMSTO) stars with distances greater
than $10$ kpc from the Galactic center, with vertical distance $\rm |Z|$ more
than $4$ kpc, and with distances less than $17.5$ kpc from the Sun, to
identify ECHOS in the inner halo. By comparison, our sample extends to four
times larger distance.

\begin{figure}
\includegraphics[width=\textwidth]{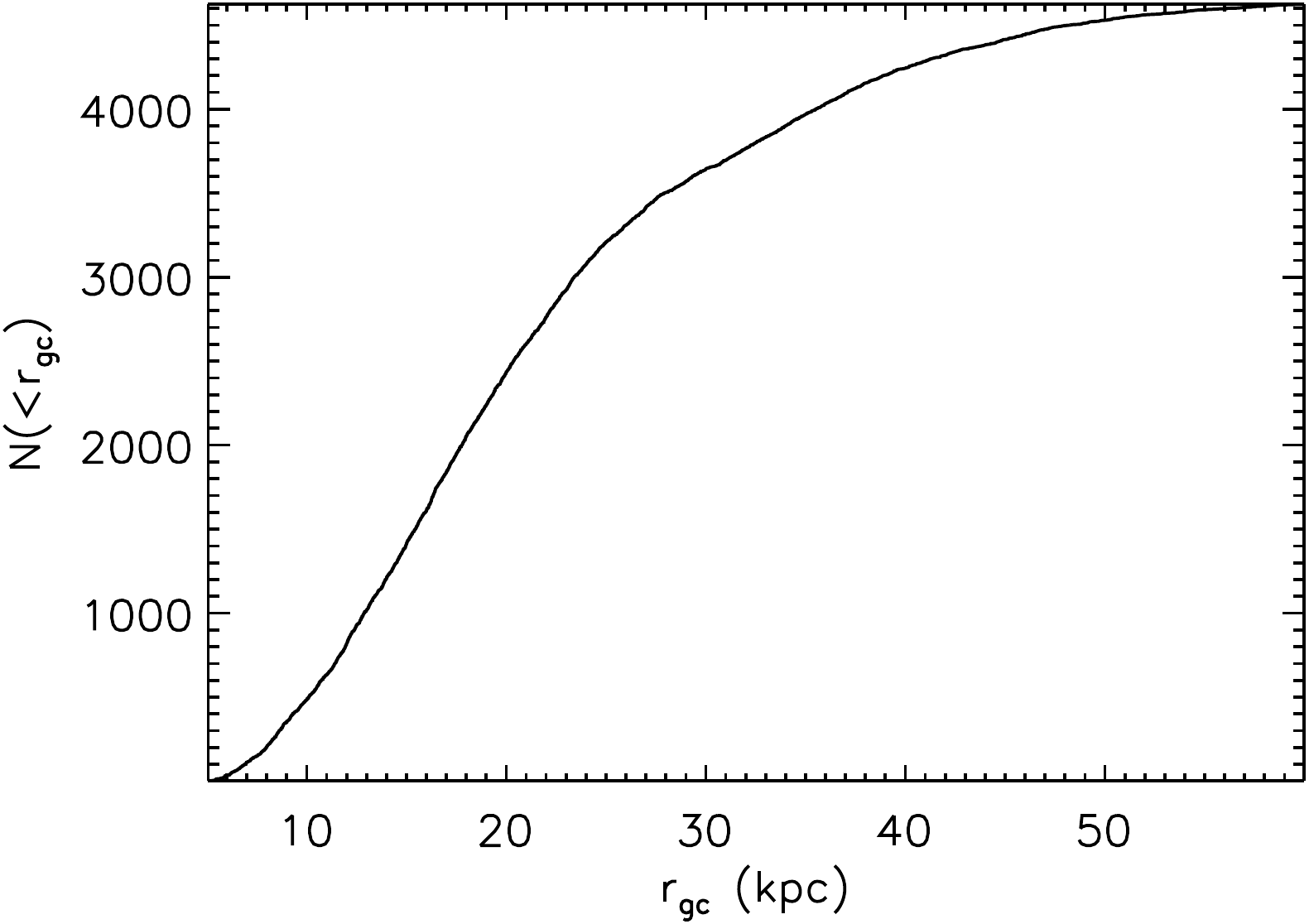}
\caption{The cumulative distribution of BHB stars with distance from the
  Galactic center, $\rm r_{gc}$, with a median distance of 22 kpc. About 90\%
  of the sample lies between 5 kpc and 40 kpc.}
\label{f:f3}
\end{figure}

The cumulative distribution of the BHB stars with $r_{gc}$ shown in
Figure~\ref{f:f3} indicates that about 95\% of the BHB stars have
$\rm{r_{gc}\leq40~kpc}$, so that any estimate of substructure should be
dominated by the BHB stars within this distance. For a cleaner selection
function, we use only the 4243 BHB stars with $\rm |Z|>4$~kpc and $\rm
r_{gc}\leq40$~kpc in the following analysis, which is still sufficiently
distant to enable tests for substructure well into the outer halo.

\section{The Close Pair Distribution of BHB stars in DR8}

We now turn to quantifying the presence of any kinematic
substructure. There is no unique choice of a substructure statistic,
nor is there a rigorous way to derive one without making very
specific assumptions about the nature of the underlying
distributions. For kinematically cold streams that are not strongly
phase-mixed, a ``pairwise velocity difference'' (PVD), $\langle |
\Delta V_{los} | \rangle (\Delta \vec{r_{gc}})$, could conceivably
be used to detect velocity substructure. It is expected that
$\langle | \Delta V_{los} | \rangle$ should be lower for small
separations $ \Delta \vec{r_{gc}}$ in stellar streams, where
adjacent stars have similar velocities. However, as many streams in
simulated halos are phase-wrapped, the PVD proved not to be very
suitable to quantify substructure, even in simulated halos where all
stars arise from disrupted satellites \citep[see][]{Xue2009}.

For distant large scale features, a great circle method \citep{LBLB1995,
  Palma2002} may be appropriate. We have chosen not to use such a method in
this paper for two reasons. Most importantly, BJ05 and \citet{Johnston2008}
demonstrate that many halo structures (in particular older structures or those
on more radial orbits) do not have a great circle geometry, and we wish to
explore the degree of halo structure in a way that is sensitive in principle
to a broader range of geometries. Secondly, the spectroscopic coverage is
sparse and covers only a fraction of the sky, and we lack useful proper motion
information for our sample, making a great circle analysis more challenging to
implement.

As an alternative to the PVD and a great circle analysis, we follow
\citet{Starkenburg2009} and \citet{dePropris2010} in exploring a statistic
that focuses on the incidence of close pairs in ($\alpha,\delta,d,V_{los}$)
space \citep[a similar approach was developed
  by][]{Doinidis1989}. Specifically, we define the separation between two
stars $i$ and $j$ as:
\begin{equation}
F_{ij}=w_{\theta}\theta_{ij}^2 +w_{\Delta d}(d_i-d_j)^2+w_{\Delta
  V_{los}}(V_{los,i}-V_{los,j})^2
\end{equation}
where
\begin{eqnarray}
\cos\theta_{ij}=\cos b_i\cos b_j\cos(l_i-l_j)+\sin b_i\sin
b_j,\nonumber\\ w_{\theta}=\frac{1}{\langle\theta^2\rangle},\nonumber w_{\Delta
  d}=\frac{1}{\langle(\Delta d)^2\rangle},\nonumber w_{\Delta V_{los}}=\frac{1}{\langle(\Delta
  V_{los})^2\rangle};\nonumber
\end{eqnarray}
\noindent and where~$\langle...\rangle$~refers to the average over all pairs.

While the angular separation combines the galactic longitude and latitude in
the distance measure F, the heliocentric distance and line-of-sight velocity
are used as independent variables... . In the definition of F, the weights
$w_{\theta}; w_{\Delta d}; w_{\Delta V_{los}}$, are solely used to create a
consistent metric for the angular, distance and velocity dimensions of F
through normalizing each dimension by the ensemble average of separation,
($\langle \theta^2 \rangle = 6928~ \rm{square degree}$), distance difference (
$\langle (\Delta d)^2 \rangle = 127~\rm{ kpc^2}$), line-of-sight velocity
difference ($\langle (\Delta V_{los})^2 \rangle = 22455~\rm{
km^2s^{-2}}$). \citet{Starkenburg2009} have pointed out that this algorithm is
quite insensitive to small changes in the weighting factors. We use somewhat
different weighting factors from them, but still detect obvious substructure
signal in the BHB sample presented here.  In parallel work to this paper
\citet{Cooper2011} applied the algorithm and weighting factors as
\citet{Starkenburg2009} to 2400 BHB stars and found obvious substructure
signal. Therefore, the choice of weighting factors seems not to be a critical
aspect of sub-structure quantification.

\begin{figure}
\centering
\includegraphics[scale=0.55]{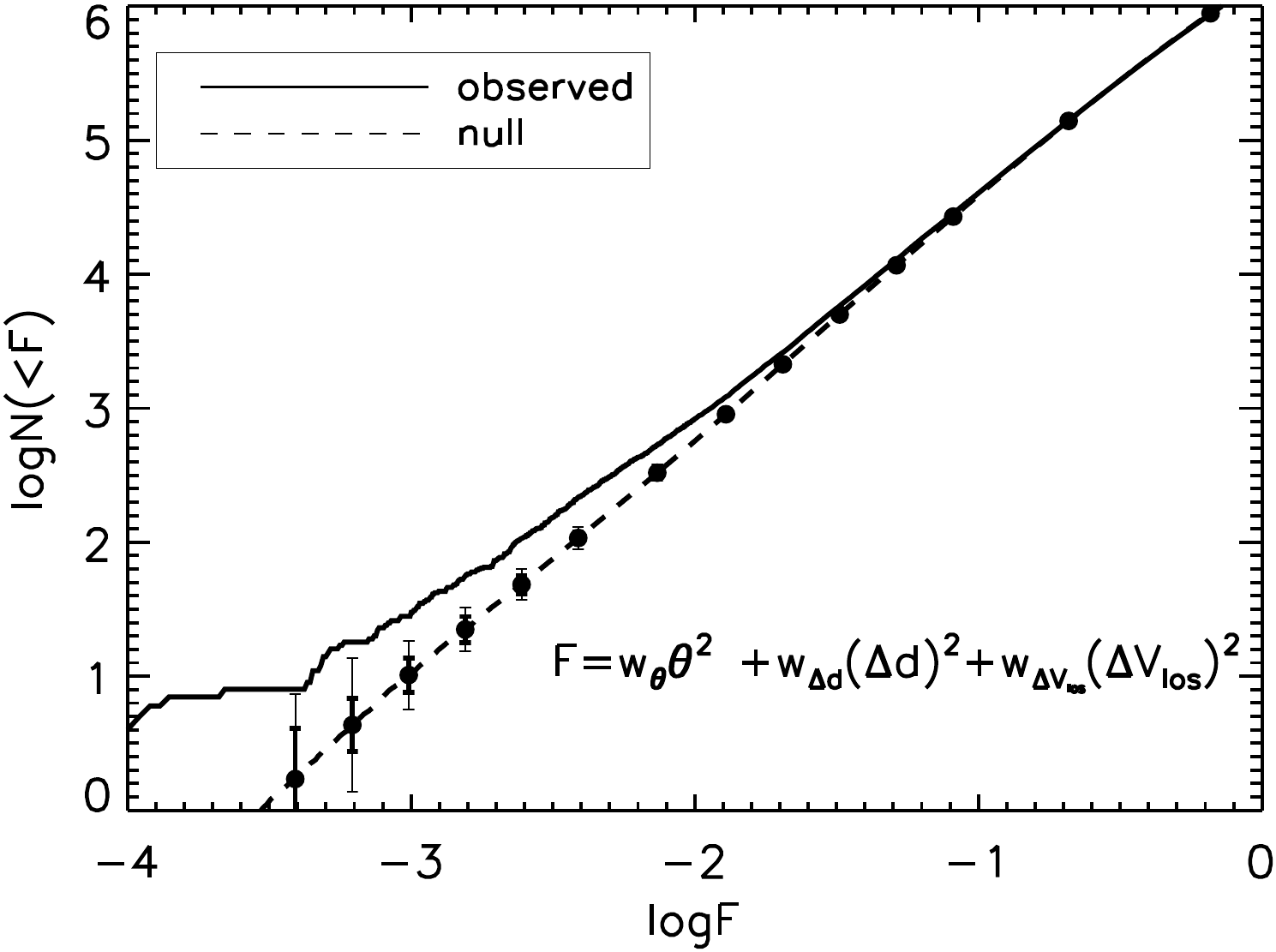}
\includegraphics[scale=0.55]{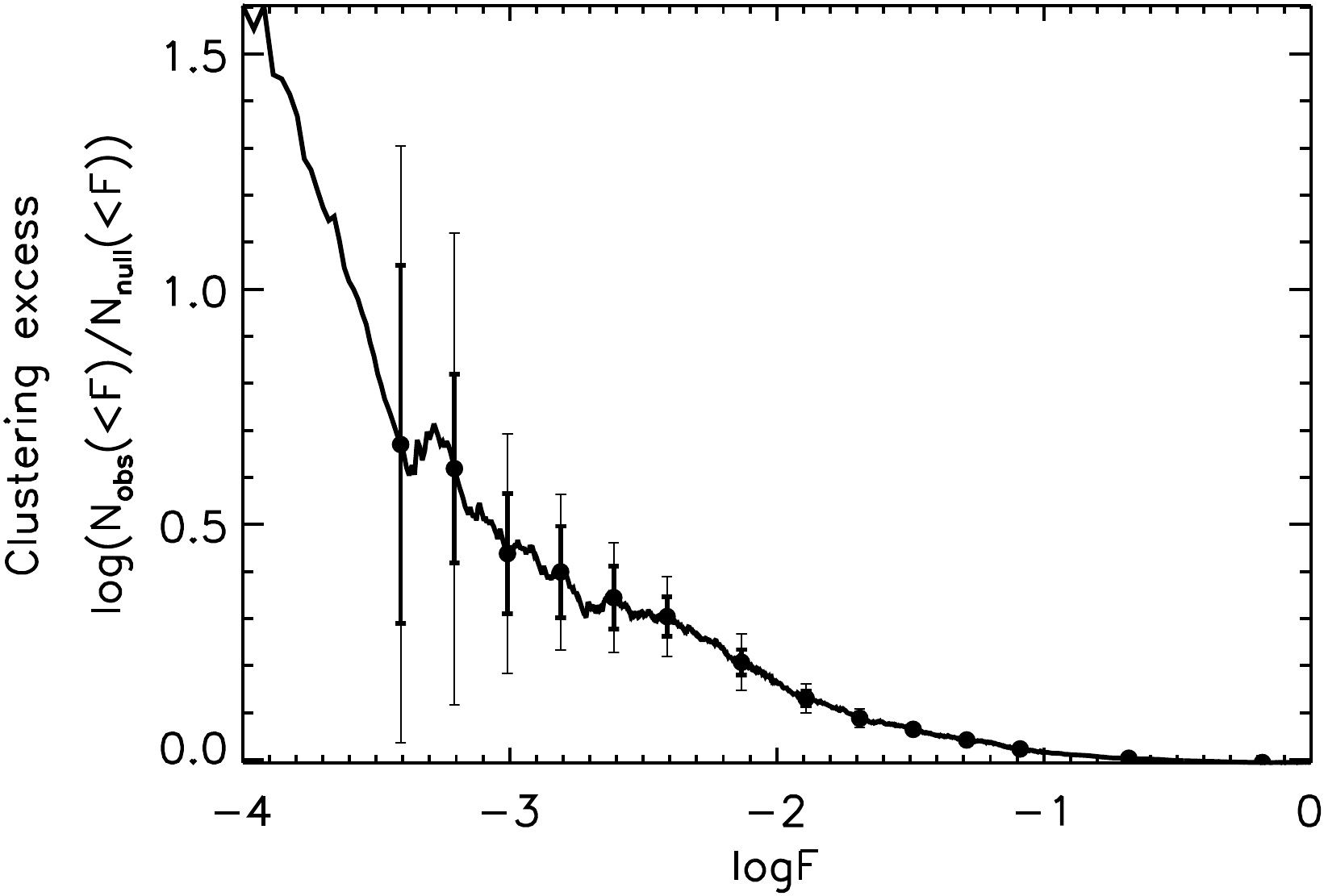}
\caption{(Upper Panel): The close pair distribution, N($<F$), for the 4243 BHB
  stars in the SDSS DR8 BHB sample with $|Z|>$ 4 kpc and $\rm r_{gc}<$ 40
  kpc. $F$ is the four-space separation between two BHB stars, taking into
  account in angle, distance, and line-of-sight velocity (Eq. 1). The solid
  line is the cumulative distribution of $F$ as observed; the dashed line is
  the average cumulative distribution of F for 100 null hypotheses, where
  positions, and hence, angular separations for each pair, were retained
  exactly as in the observations, but distances and line-of-sight velocities
  were scrambled (see Section 3). The filled circles devote the mean of 100
  such null hypotheses; the thick error bars enclose $68\%$ of the
  distribution, while the thin error bars enclose $95\%$ of the null
  hypotheses. For small F one might expect N$(< F_0)\varpropto F_0^2$ for the
  null hypothesis, but the plot shows a somewhat shallower slope, presumably
  arising from the sparse, but locally dense, angular sampling that results
  from the widely spaced SEGUE-1/2 spectroscopic plates. (Lower Panel) Ratio
  of the cumulative distribution, defined as the number of pairs in the BHB
  sample divided by the average number of null hypotheses below a certain $F$,
  N$_{obs}(<F)$/N$_{null}(<F)$. The thick and thin error bars are derived from
  those on upper plot by propagation of error. Both plots demonstrates that
  there exists a significant excess of close pairs (in distance and velocity)
  compared to the null hypothesis; BHB stars in our sample clearly exhibit
  position-velocity substructure.}
\label{f:f4}
\end{figure}

If position-velocity substructure is present, we expect that the distribution
of $F_{ij}$ for the observed sample has more close pairs than the null
hypothesis (defined below) of a smooth halo where positions and velocities are
uncorrelated: N$_{obs}(<F) >$ N$(<F_0)$. This is most conveniently captured in
the cumulative distribution of the $F_{ij}$, N$(<F)$, as illustrated in
Figure~\ref{f:f4}.

This null hypothesis assumes that the halo can be described by some spatial
density distribution, $\rho_{BHB}(\vec{r})$, and a velocity distribution where
$\sigma_{los}$ does not depend on the particular position. Indeed, averaged
over all angles, $\sigma_{los} \approx 111$ km~s$^{-1}$ is observed to be
nearly constant as a function of radius (X08). In its angular distribution and
its distance distribution, the sample selection function of our BHB sample is
very complex (see, e.g., Figure~\ref{f:f2} for the angular
distribution). However, stellar radial velocities are uncorrelated to the
sample selection, and it is reasonable to assume that the distance selection
of the stars in the same part of the sky are independent realizations of the
overall distance (or, apparent magnitude) distribution. Consequently, we
cannot randomize $\theta$ when constructing the null hypotheses. As our null
hypothesis, we can only independently draw random $\Delta d$ and $\Delta
V_{los}$.  Specifically, we do this by scrambling {\it only} the distances and
velocities within the sample to create the null hypothesis, but leave the
angular position unchanged:

\begin{equation}
F_{0,ij}=w_{\theta}\theta_{ij}^2
+w_{d}(d_{i_r}-d_{j_r})^2+w_{V_{los}}(V_{los,i_r}-V_{los,j_r})^2,
\end{equation}

\noindent where $w_{\theta}, w_{d}, w_{V_{los}}$, and the indices $i$ and $j$
are exactly the same as in $F_{ij}$, but $i_r$ and $j_r$ are random indices
chosen within an angle\footnote{Angular spacing comparable to or larger than
  the SDSS footprint may be hard to interpret, so we choose $45 \degr$ to
  avoid this.} of $45 \degr$ from stars $i$ and $j$ (here, $i_r$ and $j_r$ are
different and independent in their distance and velocity terms).

Now we can search for position-velocity substructure by comparing
N$_{obs}(<F)$ for our BHB sample to the distribution of $100$ Monte Carlo
representations of N$(<F_0)$. Figure~\ref{f:f4} shows that N$_{\rm obs}(<F)$
exceeds N$(<F_0)$ at high significance for small $F$, $\log F<-2$ (for
example, $\Delta d<1.5$~kpc, $\Delta V_{los}<15$~km~s$^{-1}$ and
$\theta<8\degr$ corresponds to $\log F<-2$). 

Small values of $F$ represent close pairs in position-velocity space. So,
Figure~\ref{f:f4} demonstrates that the observed sample has many more close
pairs than the null hypothesis, reflecting the existence of position-velocity
substructure in the BHB sample. For small F one might expect N$(<
F_0)\varpropto F_0^2$ for the null hypothesis, but the plot shows a somewhat
shallower slope, presumably arising from the non-random way that stars are
sampled by SDSS spectroscopy from the celestial sphere. The widely spaced
SEGUE-1/2 spectroscopic plates result in the sparse, but locally dense, angular
sampling. In addition, we can learn from Figure~\ref{f:f4} that the CPD
statistic focuses on $<0.1\%$ close pairs rather than all pairs of the sample,
implying that the CPD may be more sensitive to the presence of substructure
than the PVD statistic \citep[see][]{Xue2009}. 

The left panel of Figure~\ref{f:f4} shows that there are only 30 pairs with
$\log F<-3$, compared with total pairs of $8.9\times10^6$. Therefore, we just
analyze the behavior at $\log F>-3$ in the subsequent analysis.

\begin{figure}
\includegraphics[width=\textwidth]{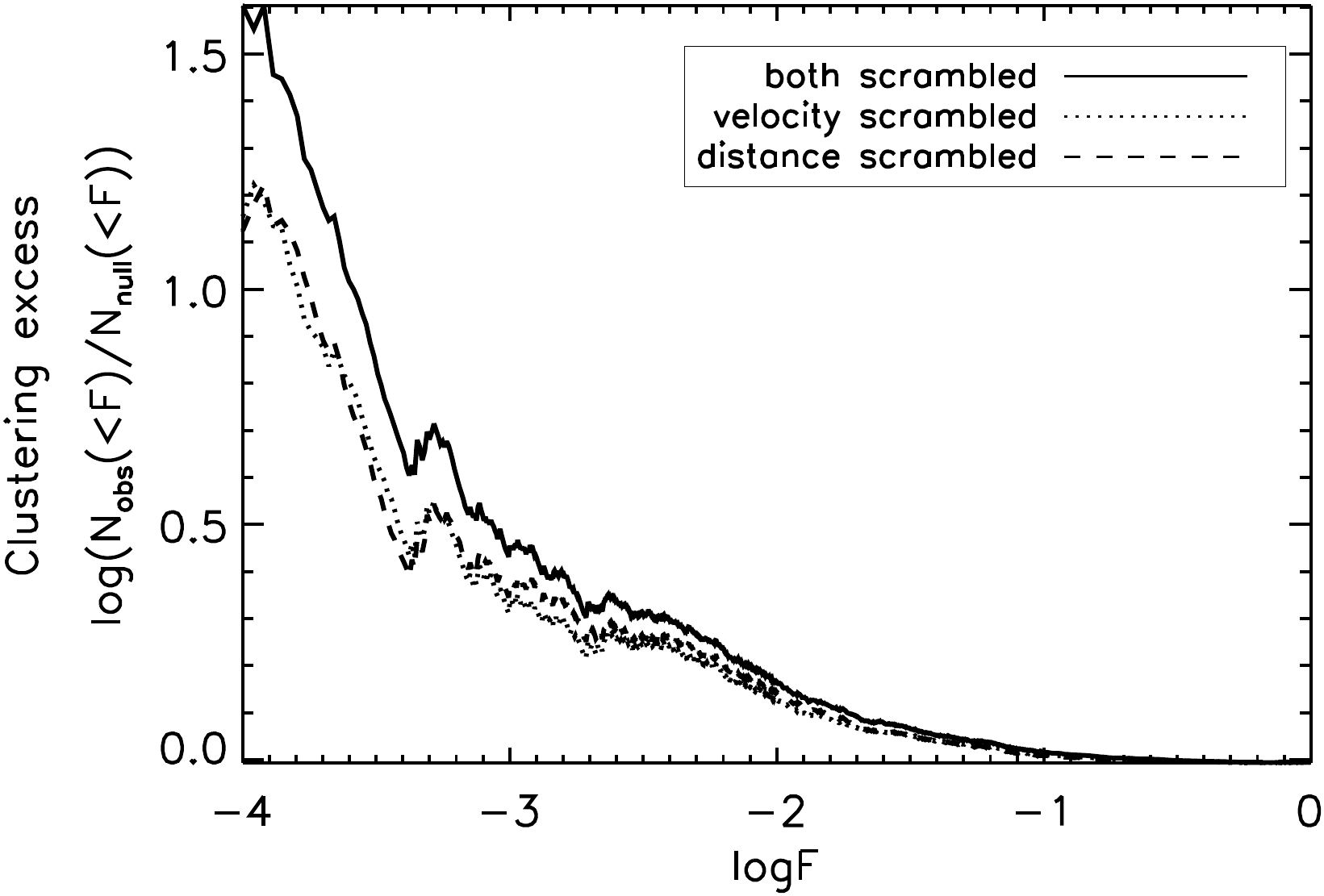}
\caption{The ratio of the cumulative distribution for the BHB sample after
  either scrambling {\it only} the distances (dashed line, case I), or {\it
    only} the velocities (dotted line, case II), or both of them (solid line,
  case III). In cases I and II an excess of close pairs is observed at a
  comparable level, but is weaker than that of case III. This implies that the
  substructure signal arises in comparable parts from both the distance and
  the line-of-sight velocity domains.}
\label{f:f5}
\end{figure}

As shown in Figure~\ref{f:f5}, the substructure signal comes both from the
distance and the line-of-sight velocity domain. This is apparent if we either
scramble {\it only} the distances or {\it only} the velocities between
N$_{obs}(<F)$ and N$(<F_0)$. In both cases an excess of small separation pairs
is present at a comparable level.

The recent studies of \citet{Carollo2007} and \citet{Carollo2010}, based on
local samples of halo stars, indicate that our Milky Way's stellar halo is
complex, and can be described by at least two components -- denoted as an
``inner'' and an ``outer'' halo, with different kinematics, distributions of
orbital eccentricity, inferred spatial profiles, and peak metallicities. In
such a decomposition, the inner halo component dominates the region of
$\rm5~kpc<r_{gc}<10$~kpc, while the region of $\rm r_{gc} > 20$~kpc is dominated
by the outer halo. Direct in situ evidence for stellar metallicity changes
with distance has also been found in photometry from the SEGUE vertical
stripes \citep{deJong2010}.

\begin{figure}
\includegraphics[width=\textwidth]{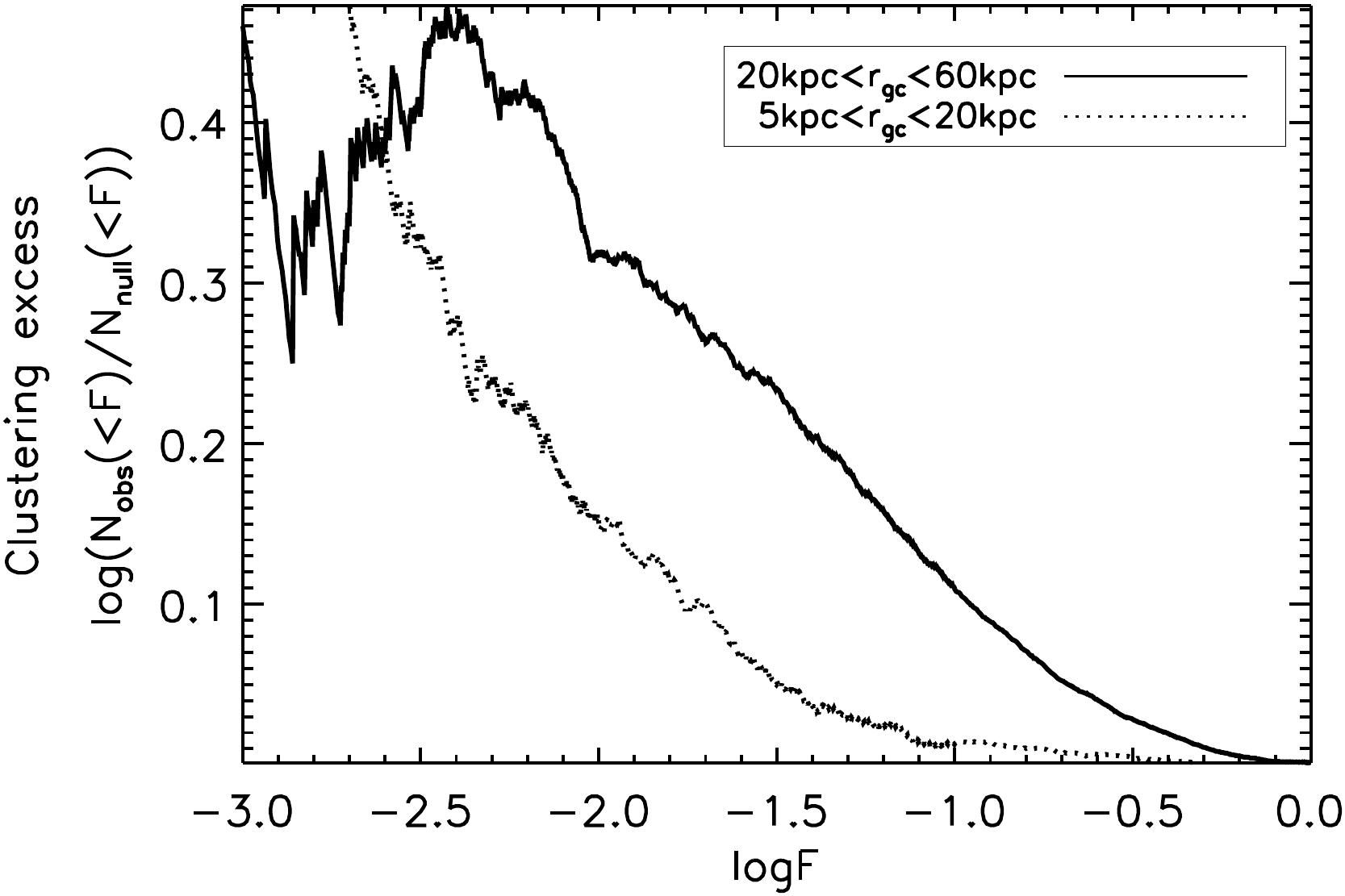}
\caption{The ratio of the cumulative distribution of $F$ for BHB stars in two
  broad Galactocentric distance ranges: subsample I, which covers 5 kpc $\rm <
  r_{gc} < $20 kpc (dotted line), subsample II, which covers 20 kpc $\rm <
  r_{gc} < $ 60 kpc (solid line). The excess of close pairs is observed in
  both subsamples. The plot illustrates that a position-velocity substructure
  signal is present in both distance ranges, covering the inner and outer
  stellar halo, and the substructure signal is more pronounced at large
  radii.}
\label{f:f6}
\end{figure}

As dynamical timescales are longer at large distances, we would expect a more
clear substructure signal in the outer parts of the halo. To test this, we
include here the BHB stars with $\rm 40~kpc<r_{gc}<60$~kpc. We divide the BHB
sample into two parts -- subsample I with $\rm5~kpc<r_{gc}<20$~kpc, and subsample
II with $\rm 20~kpc<r_{gc}<60$~kpc, both with $|Z|>4$~kpc, and compare the
substructure signals in the two subsamples. Figure~\ref{f:f6} shows that both
show significant deviations from the null hypothesis. Yet, subsample II shows
a stronger clustering excess in 4D space than subsample I. This suggests that
the substructure is stronger (e.g. less phase-mixed) in the outer halo
(subsample II) than in the inner halo (subsample I).

As mentioned in the introduction, we are more interested in a general
statistical measure of substructure for quantitative data-model comparison
than in the search for substructure associated with a particular feature;
N$(<F)$ appears as a useful statistic in this context.

\section{Position-Velocity Substructure in the BJ05 Models}

Having detected a general substructure signal, we now compare this to
expectations for N($<$F) from cosmological models where the entire stellar
halo is made of disrupted satellite galaxies.

BJ05 published models for the formation of the stellar halo of the Milky Way
system, arising solely from the accretion of $\sim 100-200$ luminous satellite
galaxies in the past $\sim 12$ Gyr. They used a hybrid semi-analytic plus
N-body approach that distinguished explicitly between the evolution of
baryonic matter and dark matter in accreted satellites. For further details of
the simulations, we refer the interested reader to BJ05,
\citet{Robertson2005}, \citet{font2006}, and references therein. There are 11
simulated halos provided by the Bullock \& Johnston study, which can be
obtained from {\it http://www.astro.columbia.edu/~kvj/halos/}.  The
simulations produce a realistic stellar halo, with mass and density profiles
much like that of the Milky Way (e.g. B08), and with surviving satellites
matching the observed number counts and structural parameter distributions of
the satellite galaxies of the Milky Way. \cite{Sharma2011} pulished a code to generate a synthetic survey of the Milky Way based on BJ05 simulations. Given one or more color-magnitude bounds, a survey size, and geometry, this code can return a catalog of stars in accordance with a given model of the Milky Way. The Galaxia code will be released publicly at {\it http://galaxia.sourceforge.net}. we refer the interested reader to \cite{Sharma2011} and references therein.

To start, we assume that BHB stars are representative tracers of the overall
population of old, metal-poor, halo stars \citep[see,
  however,][]{Bell2010}. We then make ``mock-observations'' of the BJ05
simulations, analogous to those presented in Section 2 and analyzed in Section
3. In brief, we do this by accounting for the particular survey volume of SDSS
DR8, the angular separation distribution, and approximate distance
distribution of the BHB sample, accounting for the luminosity weight of the
simulated particles, and by adding observational uncertainties for distance
and velocity.

From the simulations, we obtain the particle's $3$D positions and $3$D
velocities in the Galactic standard of rest frame, luminosities $L$, and
ages. We transfer these to Galactocentric line-of-sight velocities, $V_{los}$,
and sky positions (Galactic longitude and latitude, $(l,b)$), by taking the
Sun's position as $(8.0,0.0)$ kpc. The probability of a particle being drawn
is proportional to the assigned particle luminosity. The SDSS fibers cannot be placed closer than about 55 arcsec. However, BHB stars are relatively sparse on the sky $\sim 1$ per square degree, so possible
fiber collisions play no role in the clustering analysis. E.g. fewer than
$10^{-5}$ of all possible pairwise angular separations in the sample fall
within $<1'$. We also consider the
spectroscopic sky coverage of SDSS DR8, distance limits ($\rm |Z|\ge4$ kpc,
$\rm r_{gc} \leq 40$ kpc), and the angular separation distribution of the
observations. These procedures essentially follow those used by X08.

Based on the particles with the same sky coverage as SDSS DR8 and the same
distance limits as the BHB sample, we randomly select a particle within an
angle \footnote{The $1.2\degr$ angular distance was found to be the smallest
  angle that can ensure there is at least one particle that can be accepted
  around star $i$.} of $1.2 \degr$ from each BHB star $i$ in the sample (where
$i=1...4243$). This selected particle of luminosity L is accepted with a
probability of $\leq L/L_{max}$, where $L_{max}$ is the maximum luminosity of
the simulated particles. We also convolve the distances of the
mock-observations with an error of $5\%$; the radial velocities are convolved
with a Gaussian error of $\sigma=$5 km~s$^{-1}$.

\begin{figure}
\includegraphics[width=\textwidth]{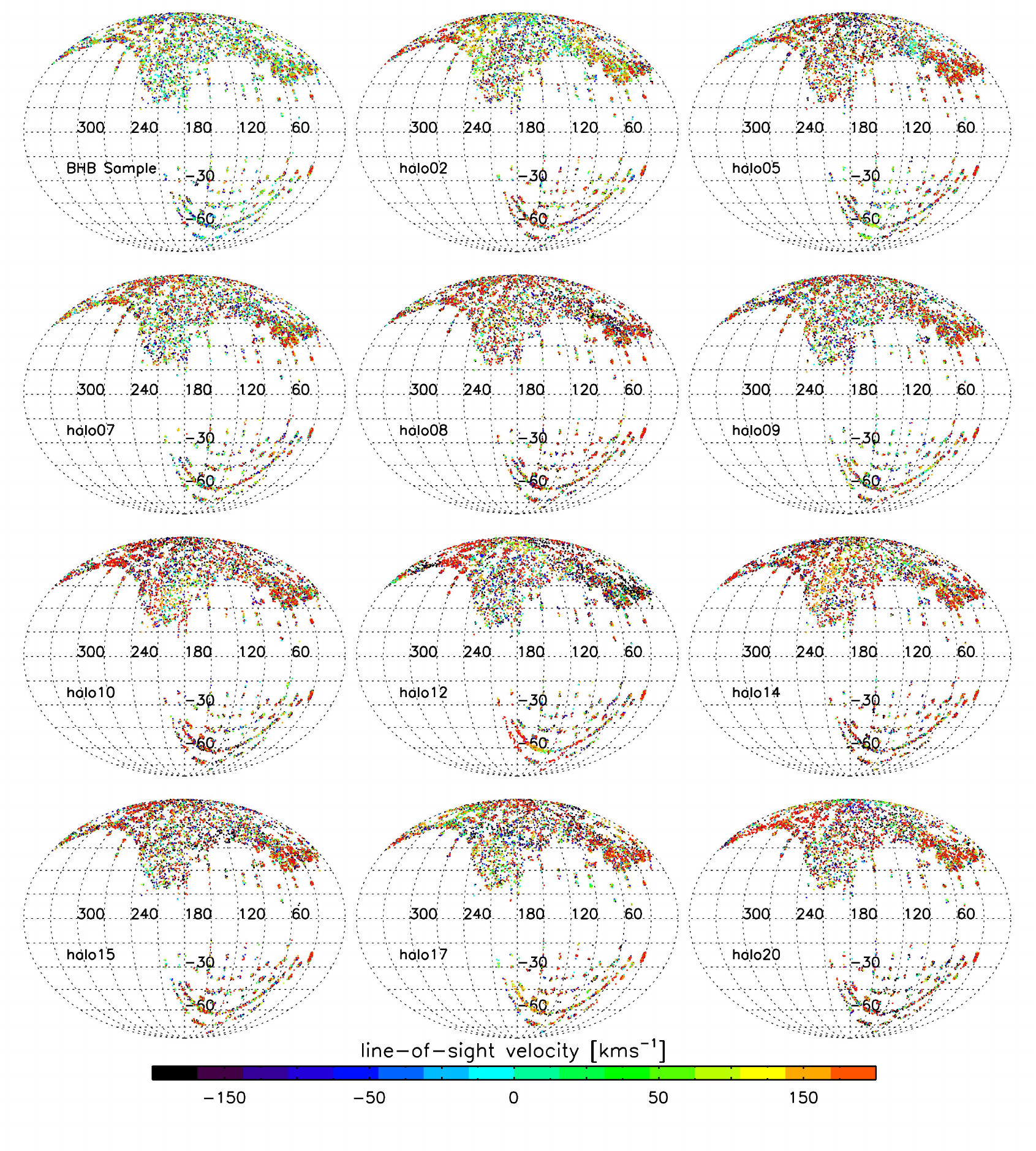}

\caption{The sky coverage for the BHB sample and 11 mock-observations from
  BJ05. The simulations were sampled in angular coverage and distance
  distribution to resemble the actual BHB sample. The stars are coded
  according to line-of-sight velocity. This figure shows that the velocity
  distributions between observation and the simulations differ somewhat. There
  are more stars with $\rm |V_{los}|>250kms^{-1}$ in the simulations.}
\label{f:f7}
\end{figure}

This procedure results in mock-observations of 4243 star particles in the
simulations that are in the same sky region as SDSS DR8, have a similar
angular separation distribution to the BHB sample, have the same distance and
velocity uncertainties as the BHB sample, and have distance limits of $\rm
|Z|\ge4$ kpc, $\rm r_{gc} \leq 40$ kpc, and satisfy the luminosity weighting
scheme. The spatial distributions for 11 mock-observations are shown in
Figure~\ref{f:f7}, along with the spatial disribution for BHB sample. These
mock-observations allow us to consider the CPD for the BJ05 simulations. We
calculate F for the mock-observations and 100 sets of the null hypothesis,
F$_0$, in each of the 11 simulations.

\begin{figure}
\centering
\includegraphics[scale=0.7]{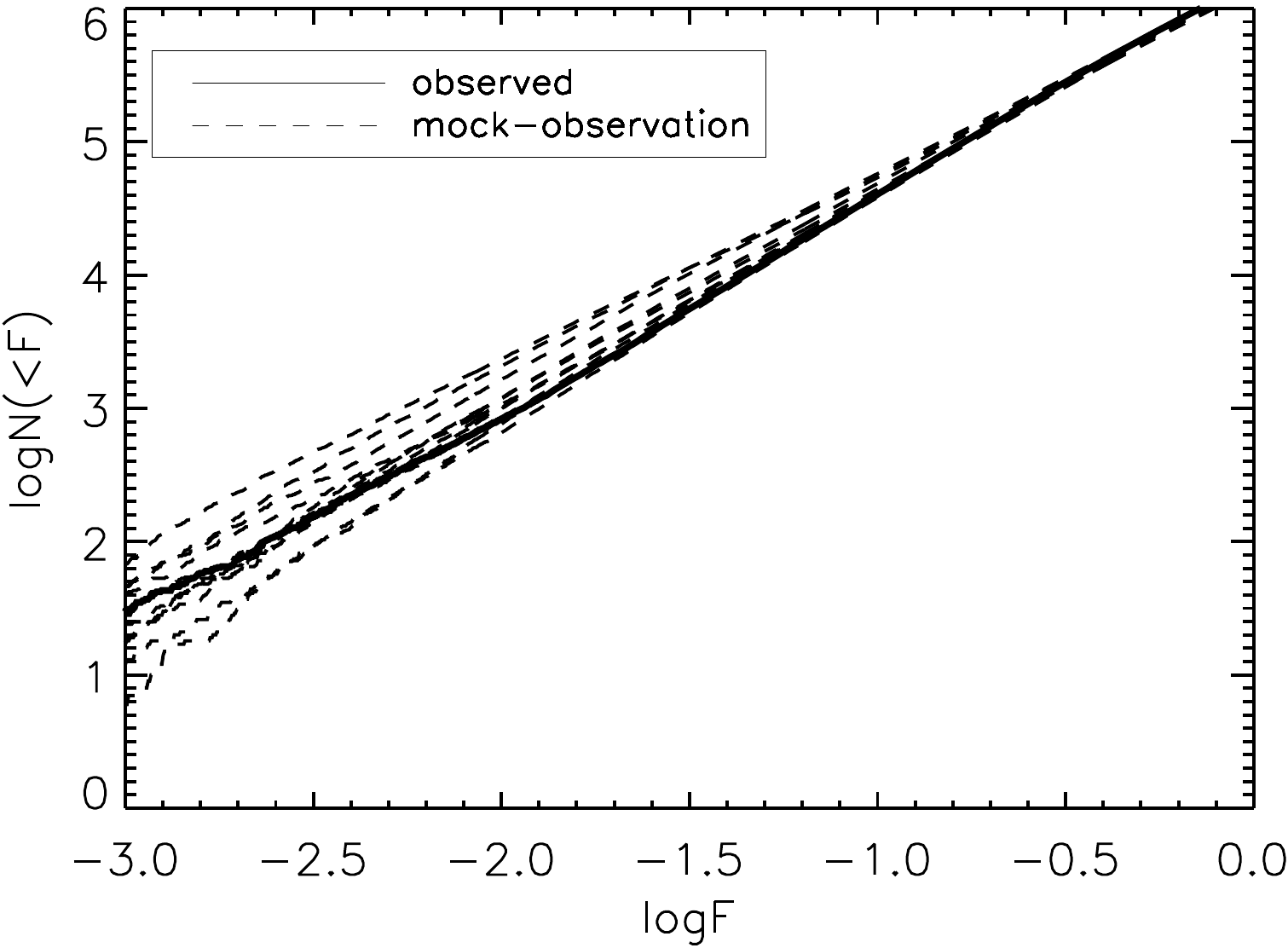}
\includegraphics[scale=0.7]{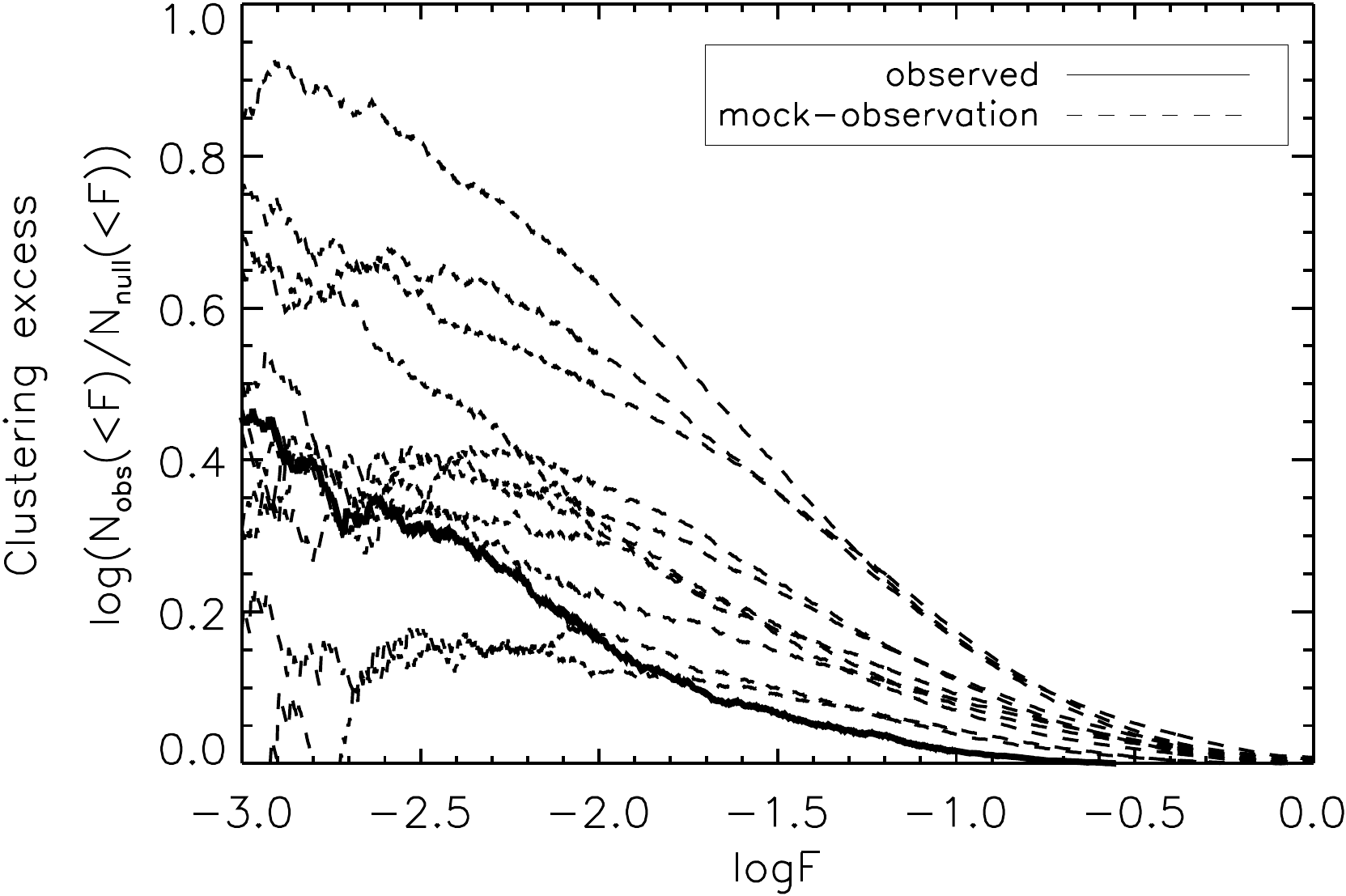}
\caption{The upper panel is the close pair distribution for the observed BHB
  sample and the 11 simulations. The solid line is the cumulative distribution
  of $F$ for the observed BHB sample; the dashed lines are the $F$
  distributions for the mock-observations of the 11 simulations. Overall, the
  observations fall well within the range of expectation from the BJ05
  simulations, but the simulations have somewhat more mid-scale power ($\log
  F~\sim~$-2.5 to -1) than the observations. The lower panel shows the
  data-model comparison for the position-velocity substructure. We show the
  ratio of the cumulative distribution for observations (solid line) and 11
  mock-observations from BJ05 (dashed lines). This figure shows that all
  simulations exhibit position-velocity clustering as an excess of
  $N_{obs}(<F)$ for small F. The observed ratio of the cumulative distribution
  is smaller than most of those seen in the simulations, where the halo is
  exclusively made up from disrupted satellites.}
\label{f:f8}
\end{figure}

The upper panel of Figure~\ref{f:f8} shows the close pair distribution for the
observed BHB sample and the 11 mock-observations. Overall, the observations
fall well within the range of expectation from the BJ05 simulations, but the
simulations have somewhat more substructure in the realm $\log F~\sim~$-2.5
to -1. The ratio of the cumulative distribution of the mock-observations for
all 11 simulated halos of BJ05 are shown in lower panel of Figure~\ref{f:f8},
along with the ratio of the cumulative disribution of the actual
data. Inspection of this Figure reveals that $N_{\rm obs}(<F)$ differs
significantly from $N(<F_0)$ for all halos, in the sense that $N_{\rm obs}(<F)
> N_{null}(<F)$ at least for $<1\%$ of closest pairs. The strength of the CPD
signature varies quite strongly among different simulations. Overall, the
ensemble of simulations show qualitatively the same signature of
position-velocity substructure as seen in the real data. Moreover, the large
majority of simulations exhibit stronger signals than observation for $-3 <
\log F < -1$ (except 2 BJ05 halos shown as lower panel of
Figure~\ref{f:f8}). In particular, a significant substructure signal can be
traced in the mock-observations to $\log F\sim-1$ (e.g., $\Delta d<4.5$~kpc,
$\Delta V_{los}<65$~km~s$^{-1}$, and $\theta<26\degr$ for $\log F<-1$) for
most of the simulations, while for the observations the substructure signal
can only be traced until $\log F\sim-2$. This indicates some quantitative
data-model difference.

The BHB sample qualitatively exhibits the same signature of position-velocity
substructure as the BJ05 halos, which are by construction 100\%
substructured. This implies that a large fraction of the Milky Way halo is
associated with substructure. However, it is difficult to quantify the
fraction of BHB stars in the Milky Way that are in substructures, because the
variation among different all-substructure simulations is so large. At any
rate, it has to be a substantial fraction.

Taken at face value, the tendency of the overall stellar halo of
cosmologically-motivated models to have exhibit more substructure than
observed BHB stars, may indicate that the models with their rigid halos may
somewhat overpredict the degree of substructure; or this comparison may imply
that some fraction of the Milky Way's halo stars did indeed form in an early
dissipational component \citep{Hammer2007,Shen2010}.

\begin{figure}
\includegraphics[width=\textwidth]{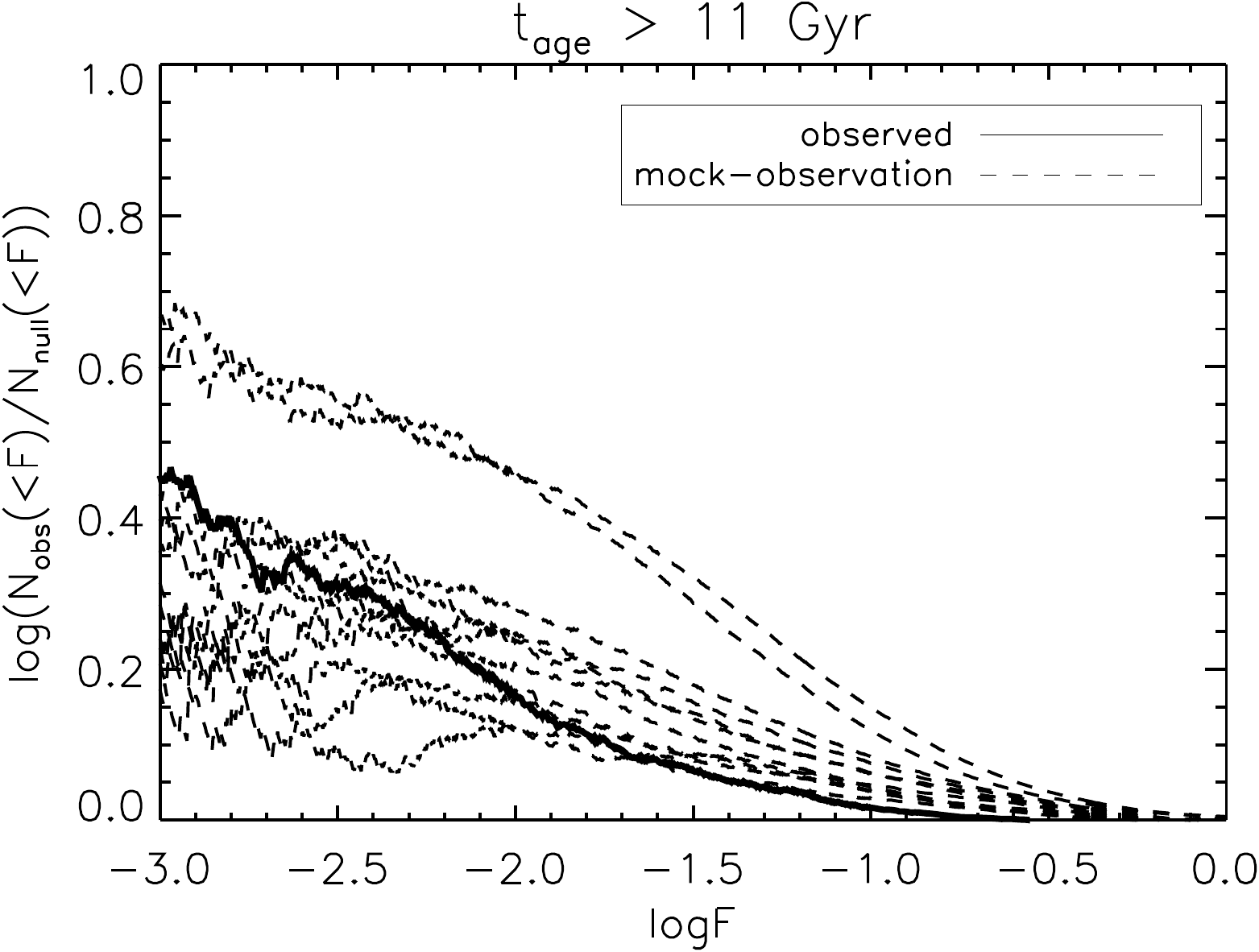}

\caption{The ratio of the cumulative distributions for the BHB sample and
  particles older than 11 Gyr in each of the 11 BJ05 simulations. The solid
  lines are the ratio of the cumulative distributions for the observation and
  the dashed lines are the ratios of the cumulative distributions for the 11
  mock-observations. Clearly, the observation is comparable to the older parts
  of most simulations (except 2 BJ05 halos).}
\label{f:f9}
\end{figure}

An alternative explanation may be this possibility, and noting that BHB stars
do not fairly trace the overall halo star populations. To test that BHB stars
occur in very old, metal poor populations, and we have tested if older star
particles from the BJ05 simulation are distributed differently from all
particles. To explore why the models might be somewhat more highly structured,
especially for larger F, we compare the $F$ distributions for the observation
and the particles older than 11 Gyr in the BJ05 halos, where the age refers to
the formation of the star particle, not the time since disruption of its host
satellite. Figure~\ref{f:f9} shows that the observed substructure signal is
comparable with those detected in simulations (Except for 2 BJ05 halos). BHB
stars are known to represent a very old population, so this may be an
astrophysically reasonable explanation for the data-model difference.

Another possibility is that, since the models do not follow mergers
self-consistently (i.e., the Milky Way's potential grows only smoothly and
analytically), the response of the Milky Way to infalling objects could serve
to disrupt and scatter streams, thereby decreasing the importance of
substructure. This could be checked in the future with simulations such as
\citet{Cooper2010}.

\begin{figure}
\includegraphics[width=\textwidth]{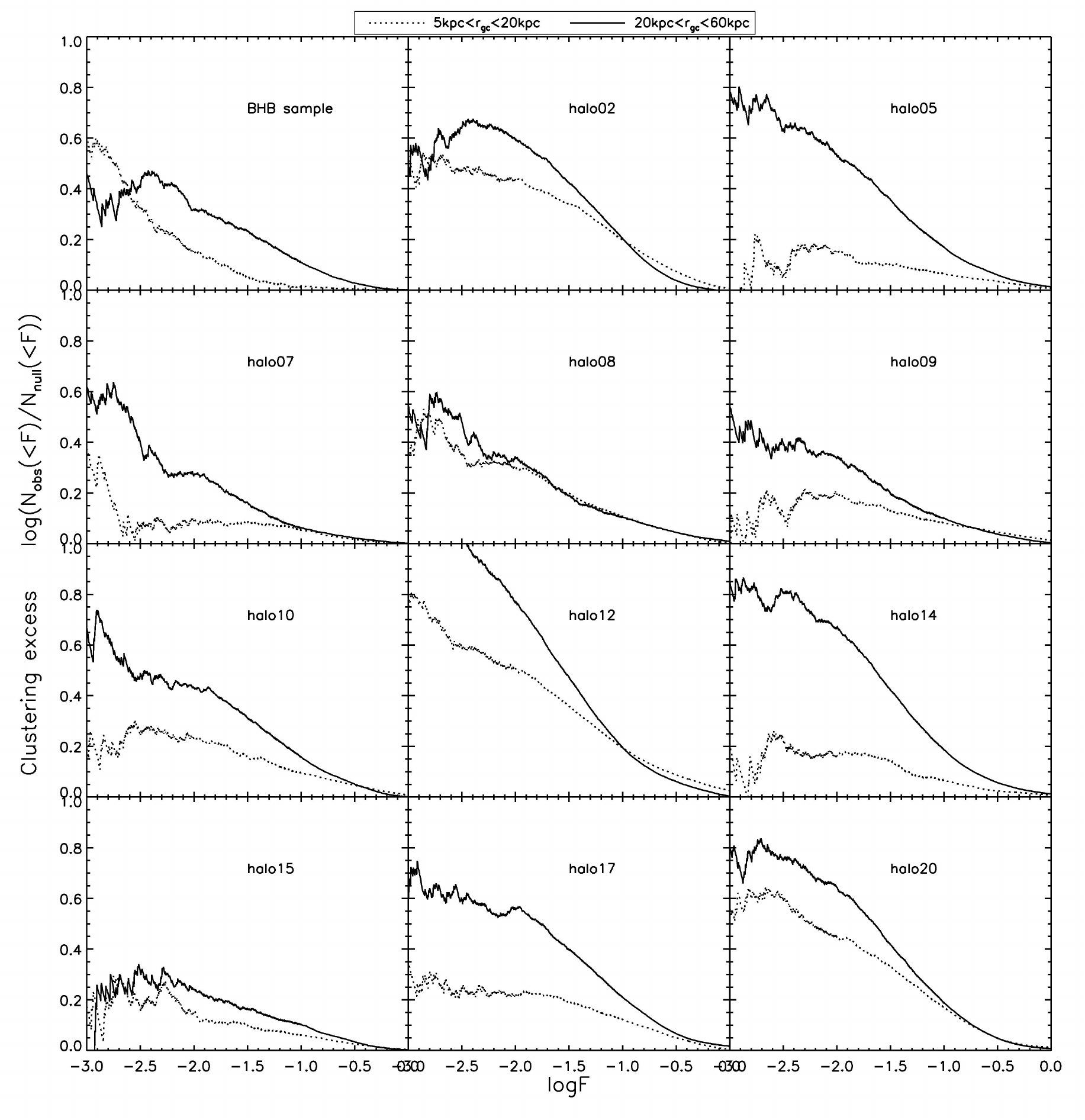}

\caption{The ratio of the cumulative distributions for BHB stars and the 11
  BJ05 simulations in regions that should be dominated by the outer-halo
  (subsample II: 20 kpc $\rm < r_{gc} < $60 kpc; solid lines) and the inner-halo
  (subsample I: 5 kpc $\rm < r_{gc} < $20 kpc; dotted lines) populations,
  respectively. The figure shows that the outer halo exhibits a more
  pronounced substructure signal than the inner halo for most simulations.}
\label{f:f10}
\end{figure}

As in the analysis of the BHB sample, we also make mock-observations of the
inner- and outer-halo regions (here, the mock-observations have similar sky
densities to the BHB sample), and calculate $F$ and $F_0$ for the
mock-observations. Figure~\ref{f:f10} shows that, for most BJ05 halos (except
halo08), the outer halo ($\rm 20~kpc<r_{gc}<60$~kpc) exhibits a stronger
substructure signal than the inner halo ($\rm 5~kpc<r_{gc}<20$~kpc),
consistent with the actual observations.

The mock-observations show that the CPD deviates from the null hypothesis in
the simulations in a qualitatively similar fashion as the actual
observations. At first sight, this seems to be a straightforward extension
into the kinematic domain of the conclusion reached by B08, that the stellar
halo exhibits a level of substructure consistent with the stream-only models
of BJ05.\footnote{In \citet{Bell2008} the BJ05 models are labeled 1-11 in
  strict numerical order, so that the interested reader can compare B08 and
  this paper side-by-side.} The signal is, however, weaker than that seen in
mock-observations drawn from the BJ05 simulations.

Taken together, Figure~\ref{f:f4}, Figure~\ref{f:f7},
Figure~\ref{f:f8}, Figure~\ref{f:f9} and Figure~\ref{f:f10} lead to
our four results: 1) In a sample of $\rm >4000$ BHB stars from SDSS
DR8 there is a very clear signal for position-velocity substructure
in the Milky Way's halo stars -- close angular pairs of stars have
smaller velocity and/or distance differences than expected for an
uncorrelated distribution. 2) The outer part of the Milky Way's halo
($\rm r_{gc}>20$~kpc) exhibits a statistically stronger kinematic
substructure signal than the inner halo ($\rm r_{gc}\leq 20$~kpc).
3) Mock-observations of simulated halos BJ05, made exclusively from
disrupted satellites, exhibit a qualitatively very similar behavior
-- N${\rm_{obs}}(<F) >$N$(<F_0)$ for $\log F\leq-1$. 4)
Quantitatively, most simulations produce a stronger signal,
especially one extending to larger scales (i.e., larger $F$).
However, if we identify BHB stars within the simulated halo population
with $t_{age}>11$ Gyr, the levels of substructure are consistent.
Given other evidence that BHB stars are most abundant in very old
populations, this seems perhaps astrophysically more plausible than
the alternative of postulating a very quiet formation of the Milky
Way's halo.


\section{Summary and Conclusions}

In the context of current cosmogonic models, the stellar halos of galaxies
like our Milky Way are expected to be comprised, to a large degree, of debris
from disrupted satellite galaxies. After disruption, the dispersing stars will
form recognizable streams for some time, but may eventually phase-mix beyond
easy recognition. There has been recent evidence (B08) that the degree of
spatial substructure actually seen in the Milky Way's halo matches that of
simulations (e.g., BJ05), where the stellar halo arises exclusively from
disrupted satellites. Due to phase-space conservation, the same scenario
qualitatively predicts the existence of a position-velocity correlation. In
this paper, we have pursued a quantitative statistical approach to
understanding how the Milky Way's stellar halo compares with this scenario.

It has already been established in the published literature that several
prominent substructures exist in the Milky Way's stellar halo, most notably
the Sagittarius stream. The next step forward is to find simple, robust
statistical measures to quantify the level of substructure in order to allow
direct comparison with theoretical models (such as BJ05). There is certainly
no established procedure, and there may be no unique way to establish such a
statistic. For example, in pure position space, B08 simply took the {\it rms}
deviation of the density from a underlying power-law model. In this paper we
have considered a statistic for diagnosing position-velocity correlations --
the close pair distribution \citep[see][]{Starkenburg2009}.

Building on recent initial attempts
\citep{Starkenburg2009,Xue2009,Harrigan2010,dePropris2010}, this paper
presented a more comprehensive attempt to quantify the position-velocity
substructure of the Milky Way's stellar halo, using BHB stars from SDSS, and
to compare it to cosmological models.  We calculated the close pair
distribution (CPD) as a function of distance separation, angular separation,
and velocity separation between pairs of stars. Qualitatively, the signal we
were looking for is that the observations have significantly more close pairs
than an ensemble of null hypotheses, where the position and velocity have no
correlation. Using this CPD (i.e., the cumulative distribution
N${\rm_{obs}}(<F)$, where $F$ is the four-distance in angle, distance, and
velocity), we found that a sample of over 4000 BHB stars in the halo of the
Galaxy exhibit far more close pairs than the null hypothesis, which
demonstrates the existence of real substructure. This result is perhaps not
surprising, as some level of substructure is already known to exist \citep[see
  also][]{Starkenburg2009,dePropris2010,Harrigan2010}. However, as a
statistical quantification, it draws on a sample 6-60 times larger than
previous analyses \citep{Starkenburg2009, dePropris2010}, and arrives at
statistically very clear-cut inferences. We also constructed mock-observations
of simulated stellar halos that are made exclusively of disrupted satellite
galaxies. These mock-observations match the angular sampling of the SDSS data
in detail, and also match the distance cuts applied to the data.  Comparing,
analogously, N${\rm_{obs}}(<F)$ to N$(<F_0)$, we found the qualitatively same
signature of substructure as in the observed sample. Quantitatively, the
observed signal is weaker than that seen in the mock-observations, where the
stellar halo is entirely made of disrupted satellites. Assuming that BHB stars
are random tracers of the stars in the simulations, we impose a lower age
limit of 11 Gyr in producing mock-observations, and found comparable levels of
position-velocity substructure between observation and simulations. Therefore,
there are two ways to reconcile the data-model differences: either to infer
differences in the dynamical formation histories between the simulated and the
observed Milky Way, or - more plausible in our view - attributing it to the
fact that BHB stars are overrepresented in the oldest sub-populations of the
stellar halo. For both the observations and the mock-observations we compared
the substructure signals associated with the inner and outer halos, and found
good agreement between data and model -- the outer halo exhibits a stronger
substructure signal than the inner halo.

Within the context of SDSS data, the next level of understanding kinematic
substructure in the Milky Way's outer halo will come from samples of more
representative giant stars with good distances. How the results from BHB stars
presented here relate to the substructure seen in main-sequence samples of the
inner halo \citep[ECHOS]{Schlaufman2009} remains to be resolved. A more recent
generation of simulations \citep[e.g.][]{Cooper2010} will also permit more far
reaching and robust conclusions.

\acknowledgments

Funding for SDSS-III has been provided by the Alfred P. Sloan Foundation, the
Participating Institutions, the National Science Foundation, and the
U.S. Department of Energy. The SDSS-III web site is http://www.sdss3.org/.

SDSS-III is managed by the Astrophysical Research Consortium for the
Participating Institutions of the SDSS-III Collaboration including the
University of Arizona, the Brazilian Participation Group, Brookhaven National
Laboratory, University of Cambridge, University of Florida, the French
Participation Group, the German Participation Group, the Instituto de
Astrofisica de Canarias, the Michigan State/Notre Dame/JINA Participation
Group, Johns Hopkins University, Lawrence Berkeley National Laboratory, Max
Planck Institute for Astrophysics, New Mexico State University, New York
University, the Ohio State University, University of Portsmouth, Princeton
University, University of Tokyo, the University of Utah, Vanderbilt
University, University of Virginia, University of Washington, and Yale
University.

This work was made possible by the support of the Max-Planck-Institute for
Astronomy, and was funded by the National Natural Science Foundation of China
(NSFC) under Nos. 10821061, 10876040 and 10973021, and supported by the Young Researcher Grant of National Astronomical Observatories, Chinese Academy of Sciences. This work was also
supported by the National Basic Research Program of China under grant
2007CB815103.

E. F. B. acknowledges NSF grant AST 1008342.

T.C.B. and Y.S.L. acknowledge partial funding of this work from grants PHY
02-16783 and PHY 08-22648: Physics Frontier Center/Joint Institute for Nuclear
Astrophysics (JINA), awarded by the U.S.  National Science Foundation.

H. Morrison acknowledges funding of this work from NSF grant AST-0098435.
\bibliographystyle{apj}
\bibliography{ref}
\clearpage

\begin{deluxetable}{rrrrrrrrrrrrrrrrrrrr}
\tablecaption{List of $4985$ BHB stars selected from SDSS DR8}
\label{t:tbl1}
\tablecolumns{20}
\rotate
\tablewidth{0pt}
\tabletypesize{\tiny}
\tablehead{\colhead{$\rm SpName$} & \colhead{$\rm RA $} & \colhead{$\rm  Dec$ }& \colhead{$\rm l$ }& \colhead{$\rm b$ }& \colhead{$\rm g$} & \colhead{$\rm u-g$} & \colhead{$\rm g-r$ }&  \colhead{$\rm D_{0.2,\delta}$} & \colhead{$\rm f_{m,\delta}$} & \colhead{$\rm c_\gamma$} & \colhead{$\rm b_\gamma$} & \colhead{d} &\colhead{r} &\colhead{x} &\colhead{y} &\colhead{z} &\colhead{HRV} & \colhead{HRVerr} & \colhead{RVgal} \\& \colhead{degree} & \colhead{degree} & \colhead{degree} & \colhead{degree} & \colhead{mag} & \colhead{mag} & \colhead{mag} & \colhead{\AA} & \colhead{$--$} & \colhead{$--$} & \colhead{\AA} &\colhead{kpc} &\colhead{kpc}&\colhead{kpc}&\colhead{kpc}&\colhead{kpc}&\colhead{$\rm km~s^{-1}$} & \colhead{$\rm km~s^{-1}$} & \colhead{$\rm km~s^{-1}$}}
\startdata
3141-55008-498 & 331.047394& 6.292428 &  66.235764 & -37.543682&  18.94&   1.19&  -0.07& 22.86&   0.25 &  0.86 &  9.17  & 46.6 &  44.7 &  -6.9 & -33.8&  -28.4& -115.8&    7.7&   46.4\\
3141-55008-477 & 330.606934 &   6.429413 &  66.000389 & -37.124928 & 16.96&   1.21&  -0.12 & 26.30 &  0.23 &  0.93  & 9.65 &  19.1  & 18.2  &  1.8&  -13.9 & -11.6& -121.4&    2.0 &  41.5\\
3141-55008-454 & 330.924957 &   7.108576 &  66.898819 & -36.909523 & 17.05 &  1.16 & -0.14 & 27.42 &  0.20 &  0.94 &  9.45 &  20.0 &  19.1  &  1.7 & -14.7 & -12.0& -275.2 &   2.3& -110.8\\
3141-55008-449&  330.812134 &   7.425377 &  67.099251 & -36.614384 & 16.05  & 1.14 & -0.17&  26.18 &  0.22 &  1.06 &  9.72&   12.6 &  12.6 &   4.1&   -9.3 &  -7.5  &  4.8  &  1.3&  170.1\\
3141-55008-432&  330.167633&    6.321596 &  65.538742 & -36.867428 & 18.35 &  1.25 & -0.14  &25.37 &  0.16 &  0.97 &  9.92&   36.4 &  34.6 &  -4.1&  -26.5 & -21.8 &-159.2 &   4.7 &   3.8\\
3141-55008-419&  330.375885 &   7.419462 &  66.734512 & -36.296391 & 16.39 &  1.21 & -0.06 & 22.91  & 0.20  & 0.90 &  8.33 &  14.4 &  14.1  &  3.4 & -10.6&   -8.5& -171.1 &   1.9&   -5.4\\
3141-55008-401 & 330.380737&    7.128237&   66.468353 & -36.493710 & 19.12 &  1.24 & -0.09 & 25.80 &  0.32 &  0.98&   9.59  & 51.8 &  49.8&   -8.6&  -38.2 & -30.8& -302.2&   10.2& -137.3\\
3141-55008-365 & 330.155487&    7.230328&   66.379021&  -36.258785&  16.51 &  1.18&  -0.05&  23.00&   0.22 &  0.83 &  7.67 &  15.2 &  14.7 &   3.1&  -11.2&   -9.0 &-185.1&    1.8 & -19.8\\
3141-55008-200 & 330.074493 &   5.726783 &  64.899216&  -37.186333&  17.73&   1.19 & -0.10 & 25.55 &  0.27 &  0.89 &  9.19 &  27.3 &  25.7 &  -1.2&  -19.7&  -16.5& -187.9&    3.3&  -26.4\\
3141-55008-097 & 330.798920 &   5.458101 &  65.236534&  -37.907768&  17.17 &  1.17 & -0.15&  27.41 &  0.24  & 1.01 & 10.44  & 21.1 &   20.0&    1.0 & -15.1 & -13.0 & -29.2 &   2.5&  131.0\\
3478-55008-051 & 241.184311  & 35.821747 &  57.356461 &  48.304852 & 16.45&   1.16 & -0.18 & 24.60  & 0.21  & 1.02 &  9.84 &  15.2 &  14.4 &   2.6 &  -8.5&   11.3& -281.2&    2.8& -146.1\\
\enddata
\tablecomments{The first column are object name composed by plate-MJD-fiberID and the next
  four columns contains the astrometry (ra, dec, l, b) for each
  object.The magnitude and color are in the next four columns:
  corrected for extinction. The next four columns are the linewidth
  parameters from the Balmer lines. The positions are listed in the
  next five columns. The radial velocities and errors are listed next.
  The complete version of this table is in the electronic
  edition of the Journal. The printed edition contains only a sample.}
\end{deluxetable}
\end{document}